\pdfoutput=1
\documentclass[11pt]{article}

\usepackage[final]{acl}

\usepackage{times}
\usepackage{latexsym}
\usepackage{enumitem}
\usepackage{multirow}
\usepackage{colortbl}
\usepackage[figuresright]{rotating}
\usepackage{longtable}
\usepackage{amssymb}
\usepackage{mathrsfs}

\newcommand{\eg}{\emph{e.g., }}

\usepackage[T1]{fontenc}
\usepackage[utf8]{inputenc}

\usepackage{microtype}

\usepackage{inconsolata}

\usepackage{graphicx}

\title{Personalized Generation In Large Model Era: A Survey}

\author{
 \textbf{Yiyan Xu\textsuperscript{1}},
 \textbf{Jinghao Zhang\textsuperscript{2}},
 \textbf{Alireza Salemi\textsuperscript{3}},
 \textbf{Xinting Hu\textsuperscript{4}\thanks{Corresponding authors.}},
\\
 \textbf{Wenjie Wang\textsuperscript{1}\footnotemark[1]},
 \textbf{Fuli Feng\textsuperscript{1}},
 \textbf{Hamed Zamani\textsuperscript{3}},
 \textbf{Xiangnan He\textsuperscript{1}},
 \textbf{Tat-Seng Chua\textsuperscript{5}}
\\
{
\fontsize{11pt}{10pt}\selectfont
 \textsuperscript{1}University of Science and Technology of China,
 \textsuperscript{2}University of Chinese Academy of Sciences
 }
 \\
{\fontsize{11pt}{10pt}\selectfont
 \textsuperscript{3}University of Massachusetts Amherst,
 \textsuperscript{4}Nanyang Technological University
}
\\
{\fontsize{11pt}{10pt}\selectfont
 \textsuperscript{5}National University of Singapore
}
\\
{\fontsize{11pt}{10pt}\selectfont
\{yiyanxu24, wenjiewang96, fulifeng93, xiangnanhe\}@gmail.com
}
\\
{\fontsize{11pt}{10pt}\selectfont
jinghao.zhang@cripac.ia.ac.cn\quad \{asalemi, zamani\}@cs.umass.edu
}
\\
{\fontsize{11pt}{10pt}\selectfont
xinting001@e.ntu.edu.sg\quad dcscts@nus.edu.sg
}
}

\begin{document}
\maketitle
\begin{abstract}
In the era of large models, content generation is gradually shifting to Personalized Generation (PGen), tailoring content to individual preferences and needs. This paper presents the first comprehensive survey on PGen, investigating existing research in this rapidly growing field. We conceptualize PGen from a unified perspective, systematically formalizing its key components, core objectives, and abstract workflows. Based on this unified perspective, we propose a multi-level taxonomy, offering an in-depth review of technical advancements, commonly used datasets, and evaluation metrics across multiple modalities, personalized contexts, and tasks. Moreover, we envision the potential applications of PGen and highlight open challenges and promising directions for future exploration. By bridging PGen research across multiple modalities, this survey serves as a valuable resource for fostering knowledge sharing and interdisciplinary collaboration, ultimately contributing to a more personalized digital landscape. 
\end{abstract}

\section{Introduction}
\label{sec:intro}
Recent advancements in large generative models have catalyzed a paradigm shift in content generation, moving from generic, one-size-fits-all generation to Personalized Generation (PGen)~\cite{wang2023generative,xu2024personalized,nguyen2024yollava}. 
By crafting personalized content that resonates more deeply with individual preferences, PGen holds great potential to enhance user-centric services and foster more engaging, immersive user experiences across various domains, such as customized product images in e-commerce~\cite{yang2024new}, personalized advertisements in marketing campaigns~\cite{tang2024genai}, and personalized AI assistants~\cite{zhang2024llm}. Given its significant potential, PGen has attracted significant attention from both academia and industry.

Despite significant progress~\cite{alaluf2025myvlm,salemi-etal-2024-lamp}, research efforts in PGen have largely evolved independently within different communities, such as Natural Language Processing (NLP), Computer Vision (CV), and Information Retrieval (IR). 
There is no survey that specifically provides a cross-community overview of PGen research. 
Existing surveys related to PGen separately follow either a model-centric or task-centric perspective, offering only partial summaries of relevant studies. 
1) Model-centric surveys focus on specific generative models for personalization, such as Multimodal Large Language Models (MLLMs)~\cite{wu2024personalized}, Large Language Models (LLMs)~\cite{zhang2024personalization,chen2024large,li2024personal,liu2025survey,li2025survey}, and Diffusion Models (DMs)~\cite{zhang2024survey}; 
2) Task-centric surveys summarize personalization techniques in specific applications, such as dialogue generation~\cite{chen-etal-2024-recent}, role-playing~\cite{chen2024persona,tseng-etal-2024-two}, and generative recommendation~\cite{ayemowa2024analysis}. 
None of these offers a unified framework that comprehensively summarizes PGen research across communities. 

\begin{figure*}[t]
% \vspace{-0.2cm}
\setlength{\abovecaptionskip}{0.05cm}
\setlength{\belowcaptionskip}{-0.45cm}
\centering
\includegraphics[scale=0.88]{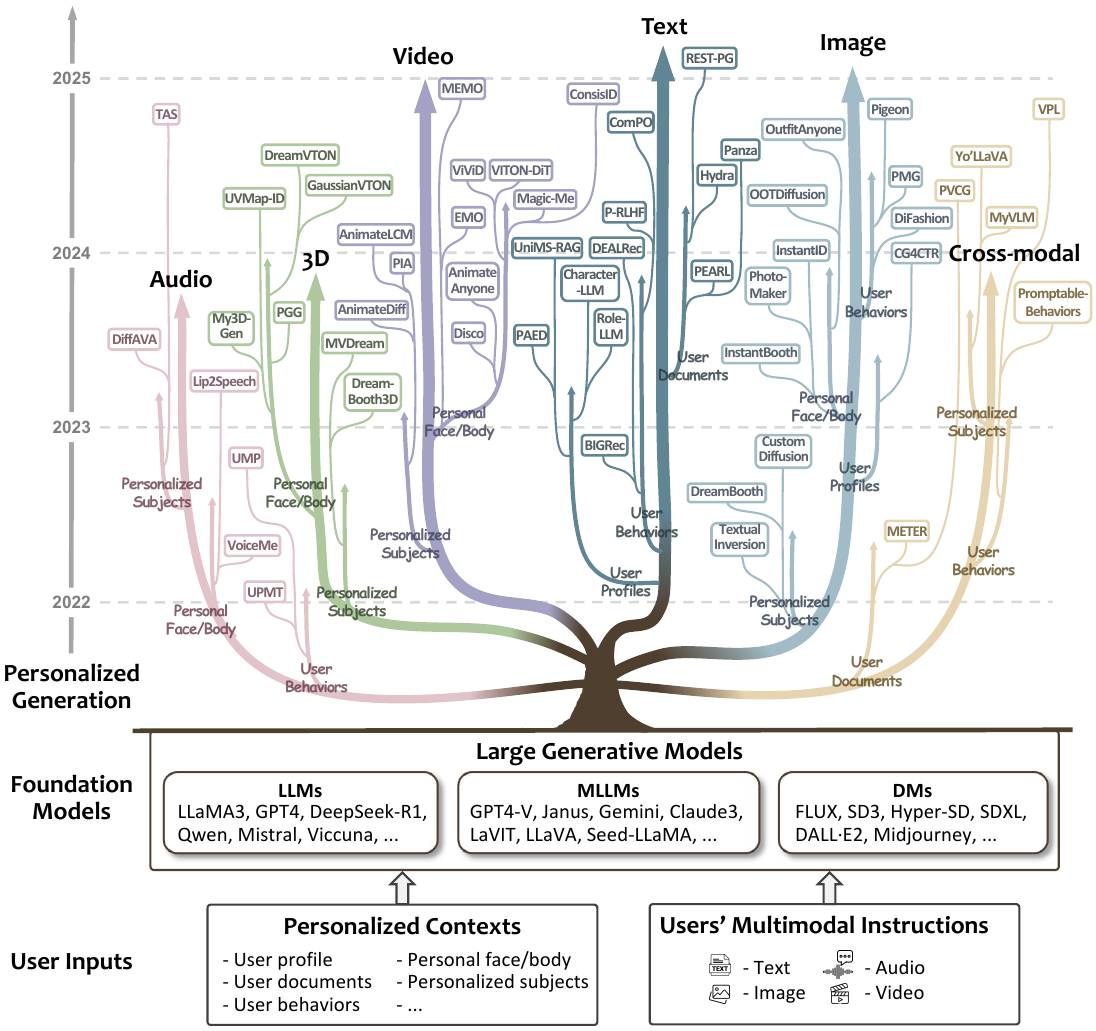}
\caption{Overview of personalized generation across modalities, inspired by the figure in~\citet{yang2024harnessing}.} 
\label{fig:tree}
\end{figure*}

A unified framework is crucial for systematically reviewing recent advances and emerging trends in PGen, providing a comprehensive, panoramic view of this field. Moreover, it can foster communication, knowledge sharing, and collaboration between various research communities, ultimately driving the development of a more advanced and personalized digital ecosystem. 
However, conducting such a unified survey is challenging, as different communities prioritize distinct modalities. For instance, the NLP and IR communities primarily focus on the text modality, while CV specializes in image, video, and 3D. Since each modality presents distinct data structures and challenges, these modality-specific differences introduce inherent technical divergences, making it difficult to unify PGen research into a cohesive framework.

\begin{figure*}[t]
% \vspace{-0.2cm}
\setlength{\abovecaptionskip}{0.05cm}
\setlength{\belowcaptionskip}{-0.45cm}
\centering
\includegraphics[scale=0.7]{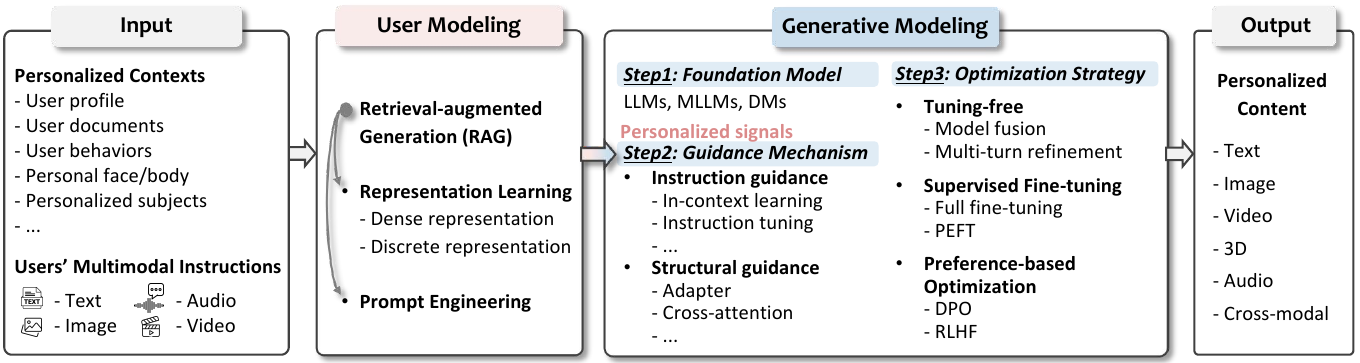}
\caption{Personalized generation workflow.} 
\label{fig:workflow}
\end{figure*}

To address these challenges, it is essential to re-examine PGen from a high-level, modality-agnostic perspective. Fundamentally, PGen entails user modeling based on various personalized contexts and multimodal instructions, extracting personalized signals to guide the content generation process. 
As illustrated in Figure~\ref{fig:tree}, existing PGen research essentially models various user inputs and leverages generative models for personalized content generation in multiple modalities.

To this end, we present the first survey dedicated to PGen. The structure of this survey and our key contributions are summarized as follows:
\begin{itemize}[leftmargin=*,itemsep=2pt,topsep=2pt,parsep=0pt]
\item \textbf{A unified user-centric perspective for PGen \textbf{(Section~\ref{sec:perspective})}.} 
We conceptualize PGen by formalizing the key components, core objectives, and general workflow, integrating studies across different modalities into a holistic framework.

\item \textbf{A multi-level taxonomy for existing work in PGen \textbf{(Section~\ref{sec:existing_work})}.} 
Building on the unified perspective, we introduce a novel taxonomy that systematically reviews PGen's technical advancements, commonly used datasets, and evaluation metrics across various modalities, personalized contexts, and tasks. 

\item \textbf{An outlook for potential applications of PGen in enhancing user-centric services \textbf{(Section~\ref{sec:application})}.} We categorize potential applications of PGen by content personalization stages, with a focus on the content creation and delivery processes.

\item \textbf{An overview of key open problems in PGen for future research \textbf{(Section \ref{sec:open_problems})}.} We outline the critical open problems that need to be addressed to drive innovation in future research and advance the user-centric content ecosystem.
\end{itemize}

\section{A Unified User-centric Perspective for Personalized Generation}
\label{sec:perspective}

\subsection{Task Formulation}
PGen leverages generative models to synthesize content tailored to individual preferences and specific needs. As illustrated in Figure~\ref{fig:tree}, it relies on two essential user inputs: 1) \textit{\textbf{Personalized contexts}} that encapsulate user preferences; 2) \textit{\textbf{Users' multimodal instructions}},which include textual prompts, voice commands, and other modality-specific inputs that explicitly convey their content needs. Generative models learn user preferences and personal characteristics from diverse personalized contexts and follow users' multimodal instructions to generate customized content across different modalities. The personalized contexts encompass the following dimensions:
\begin{itemize}[leftmargin=*,itemsep=2pt,topsep=2pt,parsep=0pt]
    \item \textit{User profiles}: A collection of demographic and personal attributes associated with a specific user, such as age, gender, occupation, and location.

    \item \textit{User documents}: User-created textual content, such as comments, emails, and social media posts, that reflects personal creative preferences.

    \item \textit{User behaviors}: User interactions captured during user engagement, such as searches, clicks, likes, comments, views, shares, and purchases.

    \item \textit{Personal face/body}: Individual facial and bodily traits, including both static features (\eg facial structure and body shape) and dynamic features (\eg expressions, gestures, and motions). These are widely used in tasks like portrait generation, fashion virtual try-ons, and 3D modeling.

    \item \textit{Personalized subjects}: User-specific concepts or entities, such as pets, personal items, and favorite objects that reflect unique tastes.
\end{itemize}
By integrating personalized contexts with users' multimodal instructions, generative models can create highly tailored content, aligning closely with individual preferences and addressing specific needs.

\subsection{Objectives}
Although PGen in each modality is shaped by unique data structures, specific challenges, and distinct tasks, three essential objectives and evaluation dimensions remain consistent across modalities:
\begin{itemize}[leftmargin=*,itemsep=2pt,topsep=2pt,parsep=0pt]
    \item \textit{\textbf{High quality}}: Ensuring the generated content meets high standards of coherence, relevance, and aesthetics.

    \item \textit{\textbf{Instruction alignment}}: Requiring the generated content to accurately adhere to users' multimodal instructions and effectively address their needs.

    \item \textit{\textbf{Personalization}}: Guaranteeing that the generated content aligns with personalized contexts and caters to specific user preferences.
\end{itemize}
While text generation has consistently achieved high-quality outputs, challenges persist in other modalities, such as image, video, audio, and 3D generation. In these domains, generated content can sometimes appear chaotic or disjointed. Maintaining high-quality standards across all modalities is fundamental to achieving successful personalized generation. 
Furthermore, in certain domains where \textit{factual accuracy} is particularly important, such as news, laws, policies, and expert knowledge, generative models must prioritize authenticity to ensure the reliability and trustworthiness of the content provided to users.

\subsection{Workflow}
As shown in Figure~\ref{fig:workflow}, the PGen workflow involves two key processes: 1) user modeling based on diverse user-specific data and 2) generative modeling across multiple modalities, ensuring high-quality, instruction-aligned, and personalized content.

\paragraph{User Modeling}
To effectively capture user preferences and specific content needs based on personalized contexts and users' multimodal instructions, three key techniques are commonly employed: 1) Representation learning, which encodes these inputs into dense embeddings~\cite{ruiz2023dreambooth,tang-etal-2024-morpheus} or summarizes them into discrete representations (\eg texts)~\cite{shen2024pmg}; 2) Prompt engineering, which involves designing task-specific prompts to structure user-specific information for generative models~\cite{chen2024tailored,10.1007/978-981-97-9431-7_17}; and 3) Retrieval-augmented generation (RAG), which enriches user-specific information by filtering out irrelevant information and integrating external relevant data~\cite{salemi2024comparingretrievalaugmentationparameterefficientfinetuning,mysore2024pearlpersonalizinglargelanguage}.
By combining these techniques, user modeling establishes a robust foundation for PGen, extracting personalized signals to guide content personalization within the generative modeling process.

\paragraph{Generative Modeling} To generate personalized content effectively, generative modeling follows a structured three-step process: 
\begin{itemize}[leftmargin=*,itemsep=2pt,topsep=2pt,parsep=0pt]
    \item \textit{\textbf{Step1: Foundation model.}} In the era of large models, LLMs, MLLMs, and DMs serve as the backbone for content generation. Selecting an appropriate foundation model based on the target modality, task requirements, and user-specific data is crucial for achieving accurate and personalized content.
    
    \item \textit{\textbf{Step2: Guidance mechanism.}} To effectively integrate personalized signals, two primary guidance mechanisms are employed: instruction guidance and structural guidance. Specifically, instruction guidance ensures models follow explicit user prompts and instructions using techniques such as in-context learning~\cite{xu2023towards,chen2024tailored,yang2023exploring} and instruction tuning~\cite{pi2024personalized,xu2024personalized}. In contrast, structural guidance modifies the model architecture by incorporating additional modules, such as adapters~\cite{ye2023ip} and cross-attention mechanisms~\cite{wei2023elite}, to embed personalized information.

    \item \textit{\textbf{Step3: Optimization Strategy.}}
    Empowering large generative models with the capability of personalized generation involves three primary optimization strategies: 1) Tuning-free methods, which utilize pre-trained models for personalized generation without modifying model parameters. These methods often rely on model fusion to assemble multiple pre-trained models~\cite{ding2024freecustom} or employ interactive generation processes that collect real-time user feedback for multi-turn refinement~\cite{von2023fabric}, ensuring alignment with individual preferences. 2) Supervised fine-tuning which optimizes model parameters using explicit supervision signals, either through full fine-tuning~\cite{xu2024diffusion,ruiz2023dreambooth} or Parameter-Efficient Fine-Tuning (PEFT) techniques~\cite{wu2024difflora,tan2024democratizing,zhang2024personalized}. 3) Preference-based optimization, which incorporates user preference data to update model parameters. A key approach is Reinforcement Learning with Human Feedback (RLHF)~\cite{li2024personalizedlanguagemodelingpersonalized,zhang2024guided}, which employs an explicit reward model to guide optimization. Alternatively, Direct Preference Optimization (DPO) offers a streamlined solution by directly aligning model parameters with pairwise user preferences~\cite{zhang2024metaalign,huang2024patchdpo}.
\end{itemize}

By integrating these advanced techniques and strategies, this workflow not only ensures adaptability to diverse personalized contexts and user instructions but also highlights the evolving landscape of large generative models, offering a scalable solution for PGen.

\section{Personalized Generation Across Modalities}
\label{sec:existing_work}
Based on the unified perspective, we present a multi-level taxonomy for PGen, systematically reviewing PGen research across various modalities, personalized contexts, and specific tasks, as outlined in Table~\ref{tab:overview}. Studies on PGen are first categorized by modality, including text, image, video, audio, 3D, and cross-modal generation. Within each modality, we further classify research based on personalized contexts and examine corresponding tasks and techniques.
Additionally, we provide a comprehensive overview of commonly used evaluation metrics and datasets for each modality, which is summarized in Table~\ref{tab:datasets}, Table~\ref{tab:metrics1}, and Table~\ref{tab:metrics2}. 

{
\renewcommand{\arraystretch}{1.1}
\begin{table*}[t]
\setlength{\abovecaptionskip}{0.05cm}
\setlength{\belowcaptionskip}{0cm}
\caption{Overview of personalized generation.}
\label{tab:overview}
\resizebox{\textwidth}{!}{
\begin{tabular}{lllp{10cm}}
\hline
\multicolumn{1}{c}{\textbf{Modality}}                            & \multicolumn{1}{c}{\textbf{Personalized Contexts}} & \multicolumn{1}{c}{\textbf{Tasks}} & \multicolumn{1}{c}{\textbf{Representative Works}} \\ \hline
\rowcolor[HTML]{A7C6D4} 
% \multicolumn{1}{c}{\cellcolor[HTML]{A7C6D4}\textbf{\begin{tabular}[c]{@{}c@{}}Text\\ (Section~\ref{sec:text})\end{tabular}}}
& \textbf{User behaviors}                            & Recommendation                     & LLM-Rec~\cite{lyu-etal-2024-llm}, DEALRec~\cite{lin2024dataefficientfinetuningllmbasedrecommendation}, BigRec~\cite{bao2023bistepgroundingparadigmlarge}, DreamRec~\cite{10.5555/3666122.3667176}               \\
\rowcolor[HTML]{A7C6D4} 
                                                                 &                                           & Information seeking                & P-RLHF~\cite{li2024personalizedlanguagemodelingpersonalized}, ComPO~\cite{kumar2024compocommunitypreferenceslanguage}                          \\ \cline{2-4} 
\rowcolor[HTML]{A7C6D4} 
                                                                 & \textbf{User documents}                            & Writing Assistant                 & REST-PG~\cite{salemi2025reasoningenhancedselftraininglongformpersonalized}, RSPG~\cite{10.1145/3626772.3657783}, Hydra~\cite{zhuang2024hydra}, PEARL~\cite{mysore2024pearlpersonalizinglargelanguage}, Panza~\cite{nicolicioiu2024panzapersonalizedtextwriting}                                      \\ \cline{2-4} 
\rowcolor[HTML]{A7C6D4} 
                                                                 & \textbf{User profiles}                             & Dialogue System                    & PAED~\cite{zhu-etal-2023-paed}, BoB~\cite{song-etal-2021-bob}, 
UniMS-RAG~\cite{wang2024unimsragunifiedmultisourceretrievalaugmented}, ORIG~\cite{chen-etal-2023-towards-robust}                        \\
\rowcolor[HTML]{A7C6D4}
\multirow[t]{-8}{*}{\cellcolor[HTML]{A7C6D4}\textbf{\begin{tabular}[c]{@{}c@{}}Text\\ (Section~\ref{sec:text})\end{tabular}}}
                                                                 &                                           & User Simulation                    & Drama Machine~\cite{magee2024dramamachinesimulatingcharacter}, Character-LLM~\cite{character-llm}, RoleLLM~\cite{rolellm}             \\ \hline
\rowcolor[HTML]{CFE2F3} 
% \multicolumn{1}{c}{\cellcolor[HTML]{CFE2F3}\textbf{\begin{tabular}[c]{@{}c@{}}Image\\ (Section~\ref{sec:image})\end{tabular}}}
& \textbf{User behaviors}                            & General-purpose generation         & PMG~\cite{shen2024pmg}, Pigeon~\cite{xu2024personalized}, PASTA~\cite{nabati2024personalized}                                \\
\rowcolor[HTML]{CFE2F3} 
                                                                 &                                           & Fashion design                     & DiFashion~\cite{xu2024diffusion}, \citet{yu2019personalized}                                    \\
\rowcolor[HTML]{CFE2F3} 
                                                                 &                                           & E-commerce product image           & AdBooster~\cite{shilova2023adbooster}, \citet{vashishtha2024chaining}, \citet{czapp2024dynamic}                               \\ \cline{2-4} 
\rowcolor[HTML]{CFE2F3} 
                                                                 & \textbf{User profiles}                             & Fashion design                     & LVA-COG~\cite{forouzandehmehr2023character}                                           \\
\rowcolor[HTML]{CFE2F3} 
                                                                 &                                           & E-commerce product image           & CG4CTR~\cite{yang2024new}                                            \\ \cline{2-4} 
\rowcolor[HTML]{CFE2F3} 
                                                                 & \textbf{Personalized subjects}                     & Subject-driven T2I generation      & Textual Inversion~\cite{gal2023an}, DreamBooth~\cite{ruiz2023dreambooth}, Custom Diffusion~\cite{kumari2023multi}   \\ \cline{2-4} 
\rowcolor[HTML]{CFE2F3} 
                                                                 & \textbf{Personal face/body}                        & Face generation                    & PhotoMaker~\cite{li2024photomaker}, InstantBooth~\cite{shi2024instantbooth}, InstantID~\cite{wang2024instantid}               \\
\rowcolor[HTML]{CFE2F3}
\multirow[t]{-11}{*}{\cellcolor[HTML]{CFE2F3}\textbf{\begin{tabular}[c]{@{}c@{}}Image\\ (Section~\ref{sec:image})\end{tabular}}}
                                                                 &                                           & Virtual try-on                     & IDM-VTON~\cite{choi2025improving}, OOTDiffusion~\cite{xu2024ootdiffusion}, OutfitAnyone~\cite{sun2024outfitanyone}              \\ \hline
\rowcolor[HTML]{E1E0FF} 
% \multicolumn{1}{c}{\cellcolor[HTML]{E1E0FF}\textbf{\begin{tabular}[c]{@{}c@{}}Video\\ (Section~\ref{sec:video})\end{tabular}}}
& \textbf{Personalized subjects}                     & Subject-driven T2V generation      & AnimateDiff~\cite{guo2024animatediff}, AnimateLCM~\cite{wang2024animatelcm}, PIA~\cite{zhang2024pia}                      \\ \cline{2-4} 
\rowcolor[HTML]{E1E0FF} 
                                                                 & \textbf{Personal face/body}                        & ID-preserving T2V generation       & Magic-Me~\cite{ma2024magic}, ID-Animator~\cite{he2024id}, ConsisID~\cite{yuan2024identity}                   \\
\rowcolor[HTML]{E1E0FF} 
                                                                 &                                           & Talking head generation            & DreamTalk~\cite{ma2023dreamtalk}, EMO~\cite{tian2025emo}, MEMO~\cite{zheng2024memo}                              \\
\rowcolor[HTML]{E1E0FF} 
                                                                 &                                           & Pose-guided video generation       & Disco~\cite{wang2023disco}, AnimateAnyone~\cite{hu2024animate}, MagicAnimate~\cite{xu2024magicanimate}                \\
\rowcolor[HTML]{E1E0FF}
\multirow[t]{-8}{*}{\cellcolor[HTML]{E1E0FF}\textbf{\begin{tabular}[c]{@{}c@{}}Video\\ (Section~\ref{sec:video})\end{tabular}}}
                                                                 &                                           & Video virtual try-on               & ViViD~\cite{fang2024vivid}, VITON-DiT~\cite{zheng2024viton}, WildVidFit~\cite{he2025wildvidfit}                      \\ \hline
\rowcolor[HTML]{EEFFE8} 
% \multicolumn{1}{c}{\cellcolor[HTML]{EEFFE8}\textbf{\begin{tabular}[c]{@{}c@{}}3D\\ (Section~\ref{sec:3D})\end{tabular}}}
& \textbf{Personalized subjects}                     & Image-to-3D generation             & MVDream~\cite{shi2023mvdream}, DreamBooth3D~\cite{raj2023dreambooth3d}, Wonder3D~\cite{long2024wonder3d}                   \\ \cline{2-4} 
\rowcolor[HTML]{EEFFE8} 
                                                                 & \textbf{Personal face/body}                        & 3D face generation                 & PoseGAN~\cite{zhang20213d}, My3DGen~\cite{qi2023my3dgen}, DiffSpeaker~\cite{ma2024diffspeaker}                     \\ \cline{2-4} 
\rowcolor[HTML]{EEFFE8} 
                                                                 &                                           & 3D human pose generation           & FewShotMotionTransfer~\cite{huang2021few}, PGG~\cite{hu2023personalized}, 3DHM~\cite{li2024synthesizing}, DreamWaltz~\cite{huang2024dreamwaltz}      \\
\rowcolor[HTML]{EEFFE8}
\multirow[t]{-6}{*}{\cellcolor[HTML]{EEFFE8}\textbf{\begin{tabular}[c]{@{}c@{}}3D\\ (Section~\ref{sec:3D})\end{tabular}}}
                                                                 &                                           & 3D virtual try-on                  & Pergamo~\cite{casado2022pergamo}, DreamVTON~\cite{xie2024dreamvton}                                \\ \hline
\rowcolor[HTML]{FFEAEE} 
% \multicolumn{1}{c}{\cellcolor[HTML]{FFEAEE}\textbf{\begin{tabular}[c]{@{}c@{}}Audio\\ (Section~\ref{sec:audio})\end{tabular}}}
& \textbf{Personal face}                             & Face-to-speech generation          & VioceMe~\cite{van2022voiceme}, FR-PSS~\cite{wang2022residual}, Lip2Speech~\cite{sheng2023zero}                       \\ \cline{2-4} 
\rowcolor[HTML]{FFEAEE} 
                                                                 & \textbf{User behaviors}                            & Music generation                   & UMP~\cite{ma2022content}, UP-Transformer~\cite{hu2022can}, UIGAN~\cite{wang2024user}                        \\ \cline{2-4} 
\rowcolor[HTML]{FFEAEE}
\multirow[t]{-4}{*}{\cellcolor[HTML]{FFEAEE}\textbf{\begin{tabular}[c]{@{}c@{}}Audio\\ (Section~\ref{sec:audio})\end{tabular}}}
                                                                 & \textbf{Personalized subjects}                     & Text-to-audio generation           & DiffAVA~\cite{mo2023diffava}, TAS~\cite{li2024tas}                                      \\ \hline
\rowcolor[HTML]{FFF8EA} 
% \multicolumn{1}{c}{\cellcolor[HTML]{FFF8EA}\textbf{\begin{tabular}[c]{@{}c@{}}Cross-Modal\\ (Section~\ref{sec:cross-modal})\end{tabular}}}
& \textbf{User behaviors}                            & Robotics                           & VPL~\cite{poddar2024personalizing}, Promptable Behaviors~\cite{hwang2024promptable}                         \\ \cline{2-4} 
\rowcolor[HTML]{FFF8EA} 
                                                                 & \textbf{User documents}                            & Caption/Comment generation         & PV-LLM~\cite{lin2024personalized}, PVCG~\cite{wu2024understanding}, METER~\cite{geng2022improving}                               \\ \cline{2-4} 
\rowcolor[HTML]{FFF8EA}
\multirow[t]{-4}{*}{\cellcolor[HTML]{FFF8EA}\textbf{\begin{tabular}[c]{@{}c@{}}Cross-Modal\\ (Section~\ref{sec:cross-modal})\end{tabular}}}
                                                                 & \textbf{Personalized subjects}                     & Cross-modal dialogue systems                 & MyVLM~\cite{alaluf2025myvlm}, Yo'LLaVA~\cite{nguyen2024yollava}, MC-LLaVA~\cite{an2024mc}                         \\ \hline
\end{tabular}
}
\end{table*}
}

{
\renewcommand{\arraystretch}{1.1}
\begin{table*}[t]
\setlength{\abovecaptionskip}{0.05cm}
\setlength{\belowcaptionskip}{0cm}
\caption{Datasets for personalized generation.}
\label{tab:datasets}
\resizebox{\textwidth}{!}{
% \begin{tabular}{lllp{10cm}}
\begin{tabular}{lllp{11cm}}
% \begin{longtable}{lllp{10cm}}
\hline
\multicolumn{1}{c}{\textbf{Modality}}                            & \multicolumn{1}{c}{\textbf{Personalized Contexts}} & \multicolumn{1}{c}{\textbf{Tasks}} & \multicolumn{1}{c}{\textbf{Datasets}}                                                                                                                                  \\ \hline

% \endfirsthead
% {{\bfseries \tablename\ \thetable{} -- continued from previous page}} \\
% \hline
% \multicolumn{1}{c}{\textbf{Modality}}                            & \multicolumn{1}{c}{\textbf{Personalized Contexts}} & \multicolumn{1}{c}{\textbf{Tasks}} & \multicolumn{1}{c}{\textbf{Datasets}} \\ \hline
% \endhead

% \hline \multicolumn{4}{|r|}{{Continued on next page}} \\ \hline
% \endfoot

% \hline \hline
% \endlastfoot

\rowcolor[HTML]{A7C6D4} 
% \multicolumn{1}{c}{\cellcolor[HTML]{A7C6D4}\textbf{\begin{tabular}[c]{@{}c@{}}Text\\ (Section~\ref{sec:text})\end{tabular}}}
& \textbf{User behaviors}                            & Recommendation                     & Amazon \cite{amazon-recom-1, amazon-recom-2}, MovieLens \cite{movie-lens}, MIND \cite{mind-new-recom}, Goodreads \cite{goodreads-1, goodreads-2}                                                                                                                                     \\
\rowcolor[HTML]{A7C6D4} 
                                                                 &                                                    & Information seeking                & SE-PQA \cite{se-pqa-search}, PWSC \cite{personalized-web-search-challenge}, AOL4PS \cite{aol4ps}                                                                                                                                                   \\ \cline{2-4} 
\rowcolor[HTML]{A7C6D4} 
                                                                 & \textbf{User documents}                            & Writing Assistant                 & LaMP \cite{salemi-etal-2024-lamp}, LongLaMP \cite{kumar2024longlampbenchmarkpersonalizedlongform}, PLAB \cite{alhafni-etal-2024-personalized}                                                                                                                                                   \\ \cline{2-4} 
\rowcolor[HTML]{A7C6D4} 
                                                                 & \textbf{User profiles}                             & Dialogue System                    & LiveChat \cite{livechat}, FoCus \cite{Jang2021CallFC}, Pchatbot \cite{10.1145/3404835.3463239}                                                                                                                                             \\
\rowcolor[HTML]{A7C6D4}
\multirow[t]{-9}{*}{\cellcolor[HTML]{A7C6D4}\textbf{\begin{tabular}[c]{@{}c@{}}Text\\ (Section~\ref{sec:text})\end{tabular}}}
                                                                 &                                                    & User Simulation                    & OpinionsQA \cite{opinionqa}, 3 RoleBench \cite{rolellm}, HPD~\cite{chen2023large}                                                                                                                                                \\ \hline
\rowcolor[HTML]{CFE2F3} 
% \multicolumn{1}{c}{\cellcolor[HTML]{CFE2F3}\textbf{\begin{tabular}[c]{@{}c@{}}Image\\ (Section~\ref{sec:image})\end{tabular}}}
& \textbf{User behaviors}                            & General-purpose generation         & Pinterest~\cite{geng2015learning}, MovieLens~\cite{movie-lens}, MIND~\cite{wu2020mind}, POG~\cite{chen2019pog}, PASTA~\cite{nabati2024personalized}, FABRIC~\cite{von2023fabric}, DialPrompt~\cite{liu2024you}, PIP~\cite{chen2024tailored}, U-sticker~\cite{chee2025106k}                                                                                                              \\
\rowcolor[HTML]{CFE2F3} 
                                                                 &                                                    & Fashion design                     & POG~\cite{chen2019pog}, Polyvore-U~\cite{lu2019learning}                                                                                                                                                        \\
\rowcolor[HTML]{CFE2F3} 
                                                                 &                                                    & E-commerce product image           & -                                                                                                                                                                      \\ \cline{2-4} 
\rowcolor[HTML]{CFE2F3} 
                                                                 & \textbf{User profiles}                             & Fashion design                     & -                                                                                                                                                                      \\
\rowcolor[HTML]{CFE2F3} 
                                                                 &                                                    & E-commerce product image           & -                                                                                                                                                                      \\ \cline{2-4} 
\rowcolor[HTML]{CFE2F3} 
                                                                 & \textbf{Personalized subjects}                     & Subject-driven T2I generation      & Dreambench~\cite{ruiz2023dreambooth}, Dreambench++~\cite{peng2024dreambench++}, CustomConcept101~\cite{kumari2023multi}, ConceptBed~\cite{patel2024conceptbed}, Textual Inverison~\cite{gal2023an}, ViCo~\cite{hao2023vico}, DreamMatcher~\cite{nam2024dreammatcher}, Break-A-Scene~\cite{avrahami2023break}, Mix-of-Show~\cite{gu2024mix}, Concept Conductor~\cite{yao2024concept}, LoRA-Composer~\cite{yang2024lora}, StyleDrop~\cite{sohn2023styledrop} \\ \cline{2-4} 
\rowcolor[HTML]{CFE2F3} 
                                                                 & \textbf{Personal face/body}                        & Face generation                    & CelebA-HQ~\cite{karras2018progressive},  FFHQ~\cite{10.1109/TPAMI.2020.2970919}, SFHQ~\cite{david_beniaguev_2022_SFHQ}, LV-MHP-v2~\cite{zhao2018understanding}, Stellar~\cite{achlioptas2023stellar}, AddMe-1.6M~\cite{yue2025addme}, FFHQ-FastComposer~\cite{xiao2024fastcomposer}, LAION-Face~\cite{zheng2022general}, PPR10K~\cite{liang2021ppr10k}, LCM-Lookahead~\cite{gal2024lcm}, CelebRef-HQ~\cite{li2022learning}, CelebV-T~\cite{yu2023celebv}, FaceForensics++~\cite{rossler2019faceforensics++}, VGGFace2~\cite{cao2018vggface2}         \\
\rowcolor[HTML]{CFE2F3}
\multirow[t]{-21}{*}{\cellcolor[HTML]{CFE2F3}\textbf{\begin{tabular}[c]{@{}c@{}}Image\\ (Section~\ref{sec:image})\end{tabular}}}
                                                                 &                                                    & Virtual try-on                     & VITON~\cite{han2018viton}, VITON-HD~\cite{choi2021viton}, DressCode~\cite{morelli2022dress}, StreetTryOn~\cite{cui2024street}, DeepFashion~\cite{ge2019deepfashion2}, Deepfashion-Multimodal~\cite{jiang2022text2human}, MPV~\cite{dong2019towards}, IGPair~\cite{shen2024imagdressing}, SHHQ~\cite{fu2022stylegan}                                                                        \\ \hline
\rowcolor[HTML]{E1E0FF} 
% \multicolumn{1}{c}{\cellcolor[HTML]{E1E0FF}\textbf{\begin{tabular}[c]{@{}c@{}}Video\\ (Section~\ref{sec:video})\end{tabular}}}
& \textbf{Personalized subjects}                     & Subject-driven T2V generation      & WebVid-10M~\cite{bain2021frozen}, UCF101~\cite{soomro2012ucf101}, AnimateBench~\cite{zhang2024pia}, VideoBooth~\cite{jiang2024videobooth}, StyleCrafter~\cite{liu2024stylecrafter}, Datasets for subject-driven T2I generation...                                                                 \\ \cline{2-4} 
\rowcolor[HTML]{E1E0FF} 
                                                                 & \textbf{Personal face/body}                        & ID-preserving T2V generation       & ID-Animator~\cite{he2024id}, ConsisID~\cite{yuan2024identity}                                                                                                                                                  \\
\rowcolor[HTML]{E1E0FF} 
                                                                 &                                                    & Talking head generation            & LRW~\cite{chung2017lip}, VoxCeleb~\cite{nagrani2020voxceleb}, VoxCeleb2~\cite{chung2018voxceleb2}, TCD-TIMIT~\cite{harte2015tcd}, LRS2~\cite{son2017lip}, HDTF~\cite{zhang2021flow}, MEAD~\cite{wang2020mead}, GRID~\cite{cooke2006audio}, MultiTalk~\cite{sungbin24_interspeech}                                                                                                 \\
\rowcolor[HTML]{E1E0FF} 
                                                                 &                                                    & Pose-guided video generation       & FashionVideo~\cite{zablotskaia2019dwnet}, TikTok~\cite{jafarian2021learning}, TED-talks~\cite{siarohin2021motion}, Everybody-dance-now~\cite{chan2019everybody}                                                                                                                   \\
\rowcolor[HTML]{E1E0FF}
\multirow[t]{-11}{*}{\cellcolor[HTML]{E1E0FF}\textbf{\begin{tabular}[c]{@{}c@{}}Video\\ (Section~\ref{sec:video})\end{tabular}}}
                                                                 &                                                    & Video virtual try-on               & VVT~\cite{dong2019fw}, ViViD~\cite{fang2024vivid}, FashionVideo~\cite{zablotskaia2019dwnet}, TikTok~\cite{jafarian2021learning}, TikTokDress~\cite{nguyen2024swifttry}                                                                                                                          \\ \hline
\rowcolor[HTML]{EEFFE8} 
% \multicolumn{1}{c}{\cellcolor[HTML]{EEFFE8}\textbf{\begin{tabular}[c]{@{}c@{}}3D\\ (Section~\ref{sec:3D})\end{tabular}}}
& \textbf{Personalized subjects}                     & Image-to-3D generation             & Dreambench~\cite{ruiz2023dreambooth}, Objaverse~\cite{deitke2023objaverse}                                                                                                                                                 \\ \cline{2-4} 
\rowcolor[HTML]{EEFFE8} 
                                                                 & \textbf{Personal face/body}                        & 3D face generation                 & Mystyle~\cite{nitzan2022mystyle}, BIWI~\cite{fanelli2013random}, VOCASET~\cite{cudeiro2019capture}                                                                                                                                                 \\
\rowcolor[HTML]{EEFFE8} 
                                                                 &                                                    & 3D human pose generation           & Human3.6M~\cite{ionescu2013human3}, 3DPW~\cite{von2018recovering}, 3DOH50K~\cite{zhang2020object}                                                                                                                                               \\
\rowcolor[HTML]{EEFFE8}
\multirow[t]{-5}{*}{\cellcolor[HTML]{EEFFE8}\textbf{\begin{tabular}[c]{@{}c@{}}3D\\ (Section~\ref{sec:3D})\end{tabular}}}
                                                                 &                                                    & 3D virtual try-on                  & BUFF~\cite{zhang2017detailed}, DreamVTON~\cite{xie2024dreamvton}                                                                                                                                                        \\ \hline
\rowcolor[HTML]{FFEAEE} 
% \multicolumn{1}{c}{\cellcolor[HTML]{FFEAEE}\textbf{\begin{tabular}[c]{@{}c@{}}Audio\\ (Section~\ref{sec:audio})\end{tabular}}}
& \textbf{Personal face}                             & Face-to-speech generation          & Voxceleb2~\cite{chung2018voxceleb2}, LibriTTS~\cite{zen2019libritts}, VGGFace2~\cite{cao2018vggface2}, GRID~\cite{cooke2006audio}, MultiTalk~\cite{sungbin24_interspeech}                                                                                                                         \\ \cline{2-4} 
\rowcolor[HTML]{FFEAEE} 
                                                                 & \textbf{User behaviors}                            & Music generation                   & Echo~\cite{Bertin-Mahieux2011}, MAESTRO~\cite{hawthorne2018enabling}                                                                                                                                                          \\ \cline{2-4} 
\rowcolor[HTML]{FFEAEE}
\multirow[t]{-4}{*}{\cellcolor[HTML]{FFEAEE}\textbf{\begin{tabular}[c]{@{}c@{}}Audio\\ (Section~\ref{sec:audio})\end{tabular}}}
                                                                 & \textbf{Personalized subjects}                     & Text-to-audio generation           & TASBench~\cite{li2024tas}, AudioCaps~\cite{kim2019audiocaps}, AudioLDM~\cite{liu2023audioldm}                                                                                                                                          \\ \hline
\rowcolor[HTML]{FFF8EA} 
% \multicolumn{1}{c}{\cellcolor[HTML]{FFF8EA}\textbf{\begin{tabular}[c]{@{}c@{}}Cross-Modal\\ (Section~\ref{sec:cross-modal})\end{tabular}}}
& \textbf{User behaviors}                            & Robotics                           & D4RL~\cite{fu2020d4rl}, Ravens~\cite{zeng2021transporter}, Habitat-Rearrange~\cite{puig2023habitat}, RoboTHOR~\cite{deitke2020robothor}                                                                                                                              \\ \cline{2-4} 
\rowcolor[HTML]{FFF8EA} 
                                                                 & \textbf{User documents}                            & Caption/Comment generation         & TripAdvisor~\cite{geng2022improving}, Yelp~\cite{geng2022improving}, PerVidCom~\cite{lin2024personalized}                                                                                                                                           \\ \cline{2-4} 
\rowcolor[HTML]{FFF8EA}
\multirow[t]{-4}{*}{\cellcolor[HTML]{FFF8EA}\textbf{\begin{tabular}[c]{@{}c@{}}Cross-Modal\\ (Section~\ref{sec:cross-modal})\end{tabular}}}
                                                                 & \textbf{Personalized subjects}                     & Cross-modal dialogue systems                 & Yo’LLaVA~\cite{nguyen2024yollava}, MyVLM~\cite{alaluf2025myvlm}                                                                                                                                               \\ \hline
\end{tabular}
% \end{longtable}
}
\end{table*}
}

% \clearpage
% \twocolumn

% \usepackage{multirow}
{
\renewcommand{\arraystretch}{1.2}
\begin{table*}[t]
\setlength{\abovecaptionskip}{0.05cm}
\setlength{\belowcaptionskip}{0cm}
\caption{Evaluation metrics for personalized text and image generation.}
\label{tab:metrics1}
\resizebox{\textwidth}{!}{
\begin{tabular}{l|p{5cm}|p{0.3cm}|p{0.3cm}|p{0.3cm}|p{0.3cm}p{0.3cm}p{0.3cm}|lp{9cm}}
\hline
\multicolumn{1}{c|}{}                                                                                                                                                                                                                                   &                                                                             & \multicolumn{1}{c|}{}                             & \multicolumn{1}{c|}{}                             & \multicolumn{1}{c|}{}                             & \multicolumn{1}{c|}{}                             & \multicolumn{1}{c|}{}                             & \multicolumn{1}{c|}{}                             &                                                  &                                                        \\
\multicolumn{1}{c|}{\multirow{-2}{*}{\textbf{Text (Section~\ref{sec:text})}}}                                                                                                                                                                                                    & \multirow{-2}{*}{\textbf{Metrics}}                                          & \multicolumn{1}{c|}{\multirow{-2}{*}{\textbf{1}}} & \multicolumn{1}{c|}{\multirow{-2}{*}{\textbf{2}}} & \multicolumn{1}{c|}{\multirow{-2}{*}{\textbf{3}}} & \multicolumn{1}{c|}{\multirow{-2}{*}{\textbf{4}}} & \multicolumn{1}{c|}{\multirow{-2}{*}{\textbf{5}}} & \multicolumn{1}{c|}{\multirow{-2}{*}{\textbf{6}}} & \multirow{-2}{*}{\textbf{Evaluation Dimensions}} & \multirow{-2}{*}{\textbf{Representative Works}}        \\ \hline
\rowcolor[HTML]{A7C6D4} 
\cellcolor[HTML]{A7C6D4}                                                                                                                                                                                                                                & NDCG  \cite{10.1145/582415.582418}                                                                      & \multicolumn{1}{c|}{\cellcolor[HTML]{A7C6D4}$\checkmark$}    & \multicolumn{1}{c|}{\cellcolor[HTML]{A7C6D4}$\checkmark$}    &                                                   & \multicolumn{1}{l|}{\cellcolor[HTML]{A7C6D4}}     & \multicolumn{1}{l|}{\cellcolor[HTML]{A7C6D4}}     &                                                   & Overall                                          & BIGRec \cite{bao2023bistepgroundingparadigmlarge}, DEALRec \cite{lin2024dataefficientfinetuningllmbasedrecommendation}, AOL4PS \cite{aol4ps}                                \\ \cline{2-10} 
\rowcolor[HTML]{A7C6D4} 
\cellcolor[HTML]{A7C6D4}                                                                                                                                                                                                                                & Hit Rate                                                                    & \multicolumn{1}{c|}{\cellcolor[HTML]{A7C6D4}$\checkmark$}    & \multicolumn{1}{c|}{\cellcolor[HTML]{A7C6D4}$\checkmark$}    &                                                   & \multicolumn{1}{l|}{\cellcolor[HTML]{A7C6D4}}     & \multicolumn{1}{l|}{\cellcolor[HTML]{A7C6D4}}     &                                                   & Overall                                          & BIGRec \cite{bao2023bistepgroundingparadigmlarge}                                                \\ \cline{2-10} 
\rowcolor[HTML]{A7C6D4} 
\cellcolor[HTML]{A7C6D4}                                                                                                                                                                                                                                & Precision                                                                   & \multicolumn{1}{c|}{\cellcolor[HTML]{A7C6D4}$\checkmark$}    & \multicolumn{1}{c|}{\cellcolor[HTML]{A7C6D4}$\checkmark$}    &                                                   & \multicolumn{1}{l|}{\cellcolor[HTML]{A7C6D4}}     & \multicolumn{1}{l|}{\cellcolor[HTML]{A7C6D4}}     &                                                   & Overall                                          & LLM-Rec \cite{lyu-etal-2024-llm}, AOL4PS \cite{aol4ps}                                       \\ \cline{2-10} 
\rowcolor[HTML]{A7C6D4} 
\cellcolor[HTML]{A7C6D4}                                                                                                                                                                                                                                & Recall                                                                      & \multicolumn{1}{c|}{\cellcolor[HTML]{A7C6D4}$\checkmark$}    & \multicolumn{1}{c|}{\cellcolor[HTML]{A7C6D4}$\checkmark$}    &                                                   & \multicolumn{1}{l|}{\cellcolor[HTML]{A7C6D4}}     & \multicolumn{1}{l|}{\cellcolor[HTML]{A7C6D4}}     &                                                   & Overall                                          & LLM-Rec \cite{lyu-etal-2024-llm}, DEALRec \cite{lin2024dataefficientfinetuningllmbasedrecommendation}, AOL4PS \cite{aol4ps}                              \\ \cline{2-10} 
\rowcolor[HTML]{A7C6D4} 
\cellcolor[HTML]{A7C6D4}                                                                                                                                                                                                                                & win-rate                                                                    &                                                   & \multicolumn{1}{c|}{\cellcolor[HTML]{A7C6D4}$\checkmark$}    &                                                   & \multicolumn{1}{l|}{\cellcolor[HTML]{A7C6D4}}     & \multicolumn{1}{l|}{\cellcolor[HTML]{A7C6D4}}     &                                                   & Overall                                          & Personalized RLHF \cite{li2024personalizedlanguagemodelingpersonalized}                                   \\ \cline{2-10} 
\rowcolor[HTML]{A7C6D4} 
\cellcolor[HTML]{A7C6D4}                                                                                                                                                                                                                                & ROUGE  \cite{lin-2004-rouge}                                                                     &                                                   &                                                   & \multicolumn{1}{c|}{\cellcolor[HTML]{A7C6D4}$\checkmark$}    & \multicolumn{1}{c|}{\cellcolor[HTML]{A7C6D4}$\checkmark$}    & \multicolumn{1}{c|}{\cellcolor[HTML]{A7C6D4}$\checkmark$}    & \multicolumn{1}{c|}{\cellcolor[HTML]{A7C6D4}$\checkmark$}    & Overall                                          & LaMP \cite{salemi-etal-2024-lamp}, RSPG \cite{10.1145/3626772.3657783}, Hydra  \cite{zhuang2024hydra}                                    \\ \cline{2-10} 
\rowcolor[HTML]{A7C6D4} 
\cellcolor[HTML]{A7C6D4}                                                                                                                                                                                                                                & BLEU   \cite{10.3115/1073083.1073135}                                                                     &                                                   &                                                   & \multicolumn{1}{c|}{\cellcolor[HTML]{A7C6D4}$\checkmark$}    & \multicolumn{1}{c|}{\cellcolor[HTML]{A7C6D4}$\checkmark$}    & \multicolumn{1}{c|}{\cellcolor[HTML]{A7C6D4}$\checkmark$}    & \multicolumn{1}{c|}{\cellcolor[HTML]{A7C6D4}$\checkmark$}    & Overall                                          & AuthorPred  \cite{li2023teachllmspersonalize}                                           \\ \cline{2-10} 
\rowcolor[HTML]{A7C6D4} 
\cellcolor[HTML]{A7C6D4}                                                                                                                                                                                                                                & BERTScore \cite{Zhang*2020BERTScore:}                                                                  &                                                   &                                                   & \multicolumn{1}{c|}{\cellcolor[HTML]{A7C6D4}$\checkmark$}    & \multicolumn{1}{c|}{\cellcolor[HTML]{A7C6D4}$\checkmark$}    & \multicolumn{1}{c|}{\cellcolor[HTML]{A7C6D4}$\checkmark$}    & \multicolumn{1}{c|}{\cellcolor[HTML]{A7C6D4}$\checkmark$}    & Overall                                          & LongLaMP     \cite{kumar2024longlampbenchmarkpersonalizedlongform}                                          \\ \cline{2-10} 
\rowcolor[HTML]{A7C6D4} 
\cellcolor[HTML]{A7C6D4}                                                                                                                                                                                                                                & GEMBA  \cite{kocmi2023largelanguagemodelsstateoftheart}                                                                     &                                                   &                                                   & \multicolumn{1}{c|}{\cellcolor[HTML]{A7C6D4}$\checkmark$}    & \multicolumn{1}{c|}{\cellcolor[HTML]{A7C6D4}$\checkmark$}    & \multicolumn{1}{c|}{\cellcolor[HTML]{A7C6D4}$\checkmark$}    & \multicolumn{1}{c|}{\cellcolor[HTML]{A7C6D4}$\checkmark$}    & Overall                                          & REST-PG     \cite{salemi2025reasoningenhancedselftraininglongformpersonalized}                                           \\ \cline{2-10} 
\rowcolor[HTML]{A7C6D4} 
\cellcolor[HTML]{A7C6D4}                                                                                                                                                                                                                                & G-Eval \cite{liu-etal-2023-g}                                                                     &                                                   &                                                   & \multicolumn{1}{c|}{\cellcolor[HTML]{A7C6D4}$\checkmark$}    & \multicolumn{1}{c|}{\cellcolor[HTML]{A7C6D4}$\checkmark$}    & \multicolumn{1}{c|}{\cellcolor[HTML]{A7C6D4}$\checkmark$}    & \multicolumn{1}{c|}{\cellcolor[HTML]{A7C6D4}$\checkmark$}    & Overall                                          & REST-PG       \cite{salemi2025reasoningenhancedselftraininglongformpersonalized}                                         \\ \cline{2-10} 
\rowcolor[HTML]{A7C6D4} 
\cellcolor[HTML]{A7C6D4}                                                                                                                                                                                                                                & ExPerT \cite{salemi2025experteffectiveexplainableevaluation}                                                                     &                                                   &                                                   & \multicolumn{1}{c|}{\cellcolor[HTML]{A7C6D4}$\checkmark$}    & \multicolumn{1}{c|}{\cellcolor[HTML]{A7C6D4}$\checkmark$}    & \multicolumn{1}{l|}{\cellcolor[HTML]{A7C6D4}}     &                                                   & Personalization                                  & ExPerT     \cite{salemi2025experteffectiveexplainableevaluation}                                            \\ \cline{2-10} 
\rowcolor[HTML]{A7C6D4} 
\cellcolor[HTML]{A7C6D4}                                                                                                                                                                                                                                & AuPEL    \cite{wang2023automatedevaluationpersonalizedtext}                                                                   &                                                   &                                                   & \multicolumn{1}{c|}{\cellcolor[HTML]{A7C6D4}$\checkmark$}    & \multicolumn{1}{c|}{\cellcolor[HTML]{A7C6D4}$\checkmark$}    & \multicolumn{1}{l|}{\cellcolor[HTML]{A7C6D4}}     &                                                   & Personalization                                  & AuPEL  \cite{wang2023automatedevaluationpersonalizedtext}                                                \\ \cline{2-10} 
\rowcolor[HTML]{A7C6D4} 
\multirow[t]{-13}{*}{\cellcolor[HTML]{A7C6D4}\textbf{\begin{tabular}[c]{@{}l@{}}1. Recommendation\\ 2. Information Seeking\\ 3. Content Generation\\ 4. Writing Assistant\\ 5. Dialogue System\\ 6. User Simulation\end{tabular}}}                  & PERSE     \cite{wang-etal-2024-learning-personalized}                                                                  &                                                   &                                                   & \multicolumn{1}{c|}{\cellcolor[HTML]{A7C6D4}$\checkmark$}    & \multicolumn{1}{c|}{\cellcolor[HTML]{A7C6D4}$\checkmark$}    & \multicolumn{1}{l|}{\cellcolor[HTML]{A7C6D4}}     &                                                   & Personalization                                  & PERSE     \cite{wang-etal-2024-learning-personalized}                                             \\ \hline
\multicolumn{1}{c|}{}                                                                                                                                                                                                                                   &                                                                             & \multicolumn{1}{c|}{}                             & \multicolumn{1}{c|}{}                             & \multicolumn{1}{c|}{}                             & \multicolumn{1}{c|}{}                             & \multicolumn{1}{c|}{}                             & \multicolumn{1}{c|}{}                             &                                                  &                                                        \\
\multicolumn{1}{c|}{\multirow{-2}{*}{\textbf{Image (Section~\ref{sec:image})}}}                                                                                                                                                                                                   & \multirow{-2}{*}{\textbf{Metrics}}                                          & \multicolumn{1}{c|}{\multirow{-2}{*}{\textbf{1}}} & \multicolumn{1}{c|}{\multirow{-2}{*}{\textbf{2}}} & \multicolumn{1}{c|}{\multirow{-2}{*}{\textbf{3}}} & \multicolumn{1}{c|}{\multirow{-2}{*}{\textbf{4}}} & \multicolumn{1}{c|}{\multirow{-2}{*}{\textbf{5}}} & \multicolumn{1}{c|}{\multirow{-2}{*}{\textbf{6}}} & \multirow{-2}{*}{\textbf{Evaluation Dimensions}} & \multirow{-2}{*}{\textbf{Representative Works}}        \\ \hline
\rowcolor[HTML]{CFE2F3} 
\cellcolor[HTML]{CFE2F3}                                                                                                                                                                                                                                & CLIP-I~\cite{radford2021learning}                                                                      & \multicolumn{1}{c|}{\cellcolor[HTML]{CFE2F3}$\checkmark$}    & \multicolumn{1}{c|}{\cellcolor[HTML]{CFE2F3}$\checkmark$}    &                                                   & \multicolumn{1}{c|}{\cellcolor[HTML]{CFE2F3}$\checkmark$}    & \multicolumn{1}{c|}{\cellcolor[HTML]{CFE2F3}$\checkmark$}    & \multicolumn{1}{c|}{\cellcolor[HTML]{CFE2F3}$\checkmark$}    & Personalization                                  & Textual Inversion~\cite{gal2023an}, Custom Diffuison~\cite{kumari2023multi}, DreamBooth~\cite{ruiz2023dreambooth}        \\ \cline{2-10} 
\rowcolor[HTML]{CFE2F3} 
\cellcolor[HTML]{CFE2F3}                                                                                                                                                                                                                                & DINO-I~\cite{caron2021emerging,oquab2024dinov}                                                                      & \multicolumn{1}{c|}{\cellcolor[HTML]{CFE2F3}$\checkmark$}    & \multicolumn{1}{c|}{\cellcolor[HTML]{CFE2F3}$\checkmark$}    &                                                   & \multicolumn{1}{c|}{\cellcolor[HTML]{CFE2F3}$\checkmark$}    & \multicolumn{1}{c|}{\cellcolor[HTML]{CFE2F3}$\checkmark$}    &                                                   & Personalization                                  & DreamBooth~\cite{ruiz2023dreambooth}, BLIP-Diffusion~\cite{li2024blip}, ELITE~\cite{wei2023elite}                      \\ \cline{2-10} 
\rowcolor[HTML]{CFE2F3} 
\cellcolor[HTML]{CFE2F3}                                                                                                                                                                                                                                & LPIPS~\cite{zhang2018unreasonable}                                                                       & \multicolumn{1}{c|}{\cellcolor[HTML]{CFE2F3}$\checkmark$}    & \multicolumn{1}{c|}{\cellcolor[HTML]{CFE2F3}$\checkmark$}    &                                                   & \multicolumn{1}{c|}{\cellcolor[HTML]{CFE2F3}$\checkmark$}    & \multicolumn{1}{l|}{\cellcolor[HTML]{CFE2F3}}     & \multicolumn{1}{c|}{\cellcolor[HTML]{CFE2F3}$\checkmark$}    & Personalization                                  & DreamSteerer~\cite{yu2024dreamsteerer}, DiFashion~\cite{xu2024diffusion}, PMG~\cite{shen2024pmg}                           \\ \cline{2-10} 
\rowcolor[HTML]{CFE2F3} 
\cellcolor[HTML]{CFE2F3}                                                                                                                                                                                                                                & PSNR~\cite{hore2010image}                                                                        &                                                   &                                                   &                                                   & \multicolumn{1}{l|}{\cellcolor[HTML]{CFE2F3}}     & \multicolumn{1}{c|}{\cellcolor[HTML]{CFE2F3}$\checkmark$}    & \multicolumn{1}{c|}{\cellcolor[HTML]{CFE2F3}$\checkmark$}    & Personalization                                  & GroupDiff~\cite{jiang2025groupdiff}, MYCloth~\cite{liu2024mycloth}, SCW-VTON~\cite{han2024shape}                           \\ \cline{2-10} 
\rowcolor[HTML]{CFE2F3} 
\cellcolor[HTML]{CFE2F3}                                                                                                                                                                                                                                & SSIM~\cite{wang2004image}                                                                        & \multicolumn{1}{c|}{\cellcolor[HTML]{CFE2F3}$\checkmark$}    & \multicolumn{1}{c|}{\cellcolor[HTML]{CFE2F3}$\checkmark$}    &                                                   & \multicolumn{1}{c|}{\cellcolor[HTML]{CFE2F3}$\checkmark$}    & \multicolumn{1}{c|}{\cellcolor[HTML]{CFE2F3}$\checkmark$}    & \multicolumn{1}{c|}{\cellcolor[HTML]{CFE2F3}$\checkmark$}    & Personalization                                  & DreamSteerer~\cite{yu2024dreamsteerer}, PMG~\cite{shen2024pmg}, OOTDifffusion~\cite{xu2024ootdiffusion}                       \\ \cline{2-10} 
\rowcolor[HTML]{CFE2F3} 
\cellcolor[HTML]{CFE2F3}                                                                                                                                                                                                                                & MS-SSIM~\cite{wang2003multiscale}                                                                     & \multicolumn{1}{c|}{\cellcolor[HTML]{CFE2F3}$\checkmark$}    & \multicolumn{1}{c|}{\cellcolor[HTML]{CFE2F3}$\checkmark$}    &                                                   & \multicolumn{1}{c|}{\cellcolor[HTML]{CFE2F3}$\checkmark$}    & \multicolumn{1}{l|}{\cellcolor[HTML]{CFE2F3}}     & \multicolumn{1}{c|}{\cellcolor[HTML]{CFE2F3}$\checkmark$}    & Personalization                                  & DreamSteerer~\cite{yu2024dreamsteerer}, Pigeon~\cite{xu2024personalized}, SieveNet~\cite{jandial2020sievenet}                         \\ \cline{2-10} 
\rowcolor[HTML]{CFE2F3} 
\cellcolor[HTML]{CFE2F3}                                                                                                                                                                                                                                & DreamSim~\cite{fu2023dreamsim}                                                                    &                                                   &                                                   &                                                   & \multicolumn{1}{c|}{\cellcolor[HTML]{CFE2F3}$\checkmark$}    & \multicolumn{1}{l|}{\cellcolor[HTML]{CFE2F3}}     & \multicolumn{1}{c|}{\cellcolor[HTML]{CFE2F3}$\checkmark$}    & Personalization                                  & IMPRINT~\cite{song2024imprint}, MaX4Zero~\cite{orzech2024masked}                                      \\ \cline{2-10} 
\rowcolor[HTML]{CFE2F3} 
\cellcolor[HTML]{CFE2F3}                                                                                                                                                                                                                                & Face similarity~\cite{deng2019arcface,schroff2015facenet,kim2022adaface,wang2018cosface}                                                             &                                                   &                                                   &                                                   & \multicolumn{1}{l|}{\cellcolor[HTML]{CFE2F3}}     & \multicolumn{1}{c|}{\cellcolor[HTML]{CFE2F3}$\checkmark$}    &                                                   & Personalization                                  & Infinite-ID~\cite{wu2025infinite}, PhotoMaker~\cite{li2024photomaker}, ProFusion~\cite{zhou2023enhancing}                     \\ \cline{2-10} 
\rowcolor[HTML]{CFE2F3} 
\cellcolor[HTML]{CFE2F3}                                                                                                                                                                                                                                & Face detection rate~\cite{deng2019arcface,zhang2016joint}                                                         &                                                   &                                                   &                                                   & \multicolumn{1}{l|}{\cellcolor[HTML]{CFE2F3}}     & \multicolumn{1}{c|}{\cellcolor[HTML]{CFE2F3}$\checkmark$}    &                                                   & Personalization                                  & SeFi-IDE~\cite{li2024sefi}, Celeb Basis~\cite{yuan2023inserting}, $\mathscr{W}_+$ Adapter~\cite{li2024stylegan}         \\ \cline{2-10} 
\rowcolor[HTML]{CFE2F3} 
\cellcolor[HTML]{CFE2F3}                                                                                                                                                                                                                                & CLIP-T~\cite{radford2021learning}                                                                      & \multicolumn{1}{c|}{\cellcolor[HTML]{CFE2F3}$\checkmark$}    & \multicolumn{1}{c|}{\cellcolor[HTML]{CFE2F3}$\checkmark$}    & \multicolumn{1}{c|}{\cellcolor[HTML]{CFE2F3}$\checkmark$}    & \multicolumn{1}{c|}{\cellcolor[HTML]{CFE2F3}$\checkmark$}    & \multicolumn{1}{c|}{\cellcolor[HTML]{CFE2F3}$\checkmark$}    &                                                   & Instruction Alignment                            & Textual Inversion~\cite{gal2023an}, Custom Diffuison~\cite{kumari2023multi}, DreamBooth~\cite{ruiz2023dreambooth}        \\ \cline{2-10} 
\rowcolor[HTML]{CFE2F3} 
\cellcolor[HTML]{CFE2F3}                                                                                                                                                                                                                                & ImageReward~\cite{xu2023imagereward}                                                                 &                                                   &                                                   &                                                   & \multicolumn{1}{c|}{\cellcolor[HTML]{CFE2F3}$\checkmark$}    & \multicolumn{1}{c|}{\cellcolor[HTML]{CFE2F3}$\checkmark$}    & \multicolumn{1}{c|}{\cellcolor[HTML]{CFE2F3}$\checkmark$}    & Instruction Alignment                            & InstructBooth~\cite{chae2023instructbooth}, DiffLoRA~\cite{wu2024difflora}, IMAGDressing-v1~\cite{shen2024imagdressing}               \\ \cline{2-10} 
\rowcolor[HTML]{CFE2F3} 
\cellcolor[HTML]{CFE2F3}                                                                                                                                                                                                                                & PickScore~\cite{kirstain2023pick}                                                                   &                                                   &                                                   &                                                   & \multicolumn{1}{c|}{\cellcolor[HTML]{CFE2F3}$\checkmark$}    & \multicolumn{1}{l|}{\cellcolor[HTML]{CFE2F3}}     &                                                   & Instruction Alignment                            & InstructBooth~\cite{chae2023instructbooth}, FABRIC~\cite{von2023fabric}, Stellar~\cite{achlioptas2023stellar}                         \\ \cline{2-10} 
\rowcolor[HTML]{CFE2F3} 
\cellcolor[HTML]{CFE2F3}                                                                                                                                                                                                                                & HPSv1~\cite{wu2023human}                                                                       &                                                   &                                                   &                                                   & \multicolumn{1}{c|}{\cellcolor[HTML]{CFE2F3}$\checkmark$}    & \multicolumn{1}{c|}{\cellcolor[HTML]{CFE2F3}$\checkmark$}    &                                                   &  Instruction Alignment                            & Stellar~\cite{achlioptas2023stellar}                                                \\ \cline{2-10} 
\rowcolor[HTML]{CFE2F3} 
\cellcolor[HTML]{CFE2F3}                                                                                                                                                                                                                                & HPSv2~\cite{wu2023human2}                                                                       &                                                   &                                                   &                                                   & \multicolumn{1}{c|}{\cellcolor[HTML]{CFE2F3}$\checkmark$}    & \multicolumn{1}{c|}{\cellcolor[HTML]{CFE2F3}$\checkmark$}    &                                                   &  Instruction Alignment                            & Stellar~\cite{achlioptas2023stellar}                                                \\ \cline{2-10} 
\rowcolor[HTML]{CFE2F3} 
\cellcolor[HTML]{CFE2F3}                                                                                                                                                                                                                                & R-precision~\cite{xu2018attngan}                                                                 &                                                   &                                                   &                                                   & \multicolumn{1}{c|}{\cellcolor[HTML]{CFE2F3}$\checkmark$}    & \multicolumn{1}{l|}{\cellcolor[HTML]{CFE2F3}}     &                                                   & Instruction Alignment                            & COTI~\cite{yang2023controllable}                                                   \\ \cline{2-10} 
\rowcolor[HTML]{CFE2F3} 
\cellcolor[HTML]{CFE2F3}                                                                                                                                                                                                                                & PAR score~\cite{gani2024llm}                                                                   &                                                   &                                                   & \multicolumn{1}{c|}{\cellcolor[HTML]{CFE2F3}$\checkmark$}    & \multicolumn{1}{l|}{\cellcolor[HTML]{CFE2F3}}     & \multicolumn{1}{l|}{\cellcolor[HTML]{CFE2F3}}     &                                                   & Instruction Alignment                            & \citet{vashishtha2024chaining}                                                    \\ \cline{2-10} 
\rowcolor[HTML]{CFE2F3} 
\cellcolor[HTML]{CFE2F3}                                                                                                                                                                                                                                & FID~\cite{heusel2017gans}                                                                         & \multicolumn{1}{c|}{\cellcolor[HTML]{CFE2F3}$\checkmark$}    & \multicolumn{1}{c|}{\cellcolor[HTML]{CFE2F3}$\checkmark$}    &                                                   & \multicolumn{1}{c|}{\cellcolor[HTML]{CFE2F3}$\checkmark$}    & \multicolumn{1}{c|}{\cellcolor[HTML]{CFE2F3}$\checkmark$}    & \multicolumn{1}{c|}{\cellcolor[HTML]{CFE2F3}$\checkmark$}    & Content Quality                                  & COTI~\cite{yang2023controllable}, IMPRINT~\cite{song2024imprint}, DiFashion~\cite{xu2024diffusion}                               \\ \cline{2-10} 
\rowcolor[HTML]{CFE2F3} 
\cellcolor[HTML]{CFE2F3}                                                                                                                                                                                                                                & KID~\cite{binkowski2018demystifying}                                                                         &                                                   &                                                   &                                                   & \multicolumn{1}{c|}{\cellcolor[HTML]{CFE2F3}$\checkmark$}    & \multicolumn{1}{c|}{\cellcolor[HTML]{CFE2F3}$\checkmark$}    & \multicolumn{1}{c|}{\cellcolor[HTML]{CFE2F3}$\checkmark$}    & Content Quality                                  & Custom Diffuison~\cite{kumari2023multi}, OOTDifffusion~\cite{xu2024ootdiffusion}, LaDI-VTON~\cite{morelli2023ladi}             \\ \cline{2-10} 
\rowcolor[HTML]{CFE2F3} 
\cellcolor[HTML]{CFE2F3}                                                                                                                                                                                                                                & IS~\cite{salimans2016improved}                                                                          &                                                   &                                                   &                                                   & \multicolumn{1}{c|}{\cellcolor[HTML]{CFE2F3}$\checkmark$}    & \multicolumn{1}{l|}{\cellcolor[HTML]{CFE2F3}}     & \multicolumn{1}{c|}{\cellcolor[HTML]{CFE2F3}$\checkmark$}    & Content Quality                                  & PE-VTON~\cite{zhang2024two}, DF-VTON~\cite{dong2024df}, Layout-and-Retouch~\cite{kim2024layout}                   \\ \cline{2-10} 
\rowcolor[HTML]{CFE2F3} 
\cellcolor[HTML]{CFE2F3}                                                                                                                                                                                                                                & LAION-Aesthetics~\cite{laion-aes}                                                            &                                                   &                                                   &                                                   & \multicolumn{1}{c|}{\cellcolor[HTML]{CFE2F3}$\checkmark$}    & \multicolumn{1}{c|}{\cellcolor[HTML]{CFE2F3}$\checkmark$}    &                                                   & Content Quality                                  & BLIP-Diffusion~\cite{li2024blip}, UniPortrait~\cite{he2024uniportrait}                            \\ \cline{2-10} 
\rowcolor[HTML]{CFE2F3} 
\cellcolor[HTML]{CFE2F3}                                                                                                                                                                                                                                & TOPIQ~\cite{chen2024topiq}                                                                       &                                                   &                                                   &                                                   & \multicolumn{1}{c|}{\cellcolor[HTML]{CFE2F3}$\checkmark$}    & \multicolumn{1}{l|}{\cellcolor[HTML]{CFE2F3}}     &                                                   & Content Quality                                  & DreamSteerer~\cite{yu2024dreamsteerer}                                           \\ \cline{2-10} 
\rowcolor[HTML]{CFE2F3} 
\cellcolor[HTML]{CFE2F3}                                                                                                                                                                                                                                & MUSIQ~\cite{ke2021musiq}                                                                       &                                                   &                                                   &                                                   & \multicolumn{1}{c|}{\cellcolor[HTML]{CFE2F3}$\checkmark$}    & \multicolumn{1}{l|}{\cellcolor[HTML]{CFE2F3}}     & \multicolumn{1}{c|}{\cellcolor[HTML]{CFE2F3}$\checkmark$}    & Content Quality                                  & DreamSteerer~\cite{yu2024dreamsteerer}, PE-VTON~\cite{zhang2024two}                                  \\ \cline{2-10} 
\rowcolor[HTML]{CFE2F3} 
\cellcolor[HTML]{CFE2F3}                                                                                                                                                                                                                                & MANIQA~\cite{yang2022maniqa}                                                                      &                                                   &                                                   &                                                   & \multicolumn{1}{l|}{\cellcolor[HTML]{CFE2F3}}     & \multicolumn{1}{l|}{\cellcolor[HTML]{CFE2F3}}     & \multicolumn{1}{c|}{\cellcolor[HTML]{CFE2F3}$\checkmark$}    & Content Quality                                  & PE-VTON~\cite{zhang2024two}                                                \\ \cline{2-10} 
\rowcolor[HTML]{CFE2F3} 
\cellcolor[HTML]{CFE2F3}                                                                                                                                                                                                                                & LIQE~\cite{zhang2023blind}                                                                        &                                                   &                                                   &                                                   & \multicolumn{1}{c|}{\cellcolor[HTML]{CFE2F3}$\checkmark$}    & \multicolumn{1}{l|}{\cellcolor[HTML]{CFE2F3}}     &                                                   & Content Quality                                  & DreamSteerer~\cite{yu2024dreamsteerer}                                           \\ \cline{2-10} 
\rowcolor[HTML]{CFE2F3} 
\cellcolor[HTML]{CFE2F3}                                                                                                                                                                                                                                & QS~\cite{gu2020giqa}                                                                          &                                                   &                                                   &                                                   & \multicolumn{1}{c|}{\cellcolor[HTML]{CFE2F3}$\checkmark$}    & \multicolumn{1}{l|}{\cellcolor[HTML]{CFE2F3}}     &                                                   & Content Quality                                  & AddMe~\cite{yue2025addme}                                                  \\ \cline{2-10} 
\rowcolor[HTML]{CFE2F3} 
\cellcolor[HTML]{CFE2F3}                                                                                                                                                                                                                                & BRISQUE~\cite{mittal2012no}                                                                     &                                                   &                                                   & \multicolumn{1}{c|}{\cellcolor[HTML]{CFE2F3}$\checkmark$}    & \multicolumn{1}{l|}{\cellcolor[HTML]{CFE2F3}}     & \multicolumn{1}{l|}{\cellcolor[HTML]{CFE2F3}}     &                                                   & Content Quality                                  & \citet{vashishtha2024chaining}                                                    \\ \cline{2-10} 
\rowcolor[HTML]{CFE2F3} 
\cellcolor[HTML]{CFE2F3}                                                                                                                                                                                                                                & CTR                                                                         &                                                   &                                                   & \multicolumn{1}{c|}{\cellcolor[HTML]{CFE2F3}$\checkmark$}    & \multicolumn{1}{l|}{\cellcolor[HTML]{CFE2F3}}     & \multicolumn{1}{l|}{\cellcolor[HTML]{CFE2F3}}     &                                                   & Overall                                          & CG4CTR~\cite{yang2024new}, \citet{czapp2024dynamic}                                            \\ \cline{2-10} 
\rowcolor[HTML]{CFE2F3} 
\cellcolor[HTML]{CFE2F3}                                                                                                                                                                                                                                & Stellar metrics~\cite{achlioptas2023stellar}                                                             &                                                   &                                                   &                                                   & \multicolumn{1}{c|}{\cellcolor[HTML]{CFE2F3}$\checkmark$}    & \multicolumn{1}{c|}{\cellcolor[HTML]{CFE2F3}$\checkmark$}    &                                                   & Overall                                          & Stellar~\cite{achlioptas2023stellar}                                                \\ \cline{2-10} 
\rowcolor[HTML]{CFE2F3} 
\multirow[t]{-42}{*}{\cellcolor[HTML]{CFE2F3}\textbf{\begin{tabular}[c]{@{}l@{}}1. General-purpose generation\\ 2. Fashion design\\ 3. E-commerce product image\\ 4. Subject-driven T2I generation\\ 5. Face generation\\ 6. Virtual try-on\end{tabular}}} & CAMI~\cite{shen2024imagdressing}                                                                        &                                                   &                                                   &                                                   & \multicolumn{1}{l|}{\cellcolor[HTML]{CFE2F3}}     & \multicolumn{1}{l|}{\cellcolor[HTML]{CFE2F3}}     & \multicolumn{1}{c|}{\cellcolor[HTML]{CFE2F3}$\checkmark$}    & Overall                                          & IMAGDressing-v1~\cite{shen2024imagdressing}                                        \\ \hline
\end{tabular}
}
\end{table*}
}

{
\renewcommand{\arraystretch}{0.94}
\begin{table*}[t]
\setlength{\abovecaptionskip}{0.05cm}
\setlength{\belowcaptionskip}{0cm}
\caption{Evaluation metrics for personalized generation across video, 3D, audio, and cross-modal domains.}
\label{tab:metrics2}
\resizebox{\textwidth}{!}{
\begin{tabular}{l|p{5cm}|p{0.3cm}|p{0.3cm}|p{0.3cm}|p{0.3cm}p{0.3cm}p{0.3cm}|lp{9cm}}
\hline
\multicolumn{1}{c|}{}                                                                                                                                                                                                                                   &                                                                             & \multicolumn{1}{c|}{}                             & \multicolumn{1}{c|}{}                             & \multicolumn{1}{c|}{}                             & \multicolumn{1}{c|}{}                             & \multicolumn{1}{c|}{}                             & \multicolumn{1}{c|}{}                             &                                                  &                                                        \\
\multicolumn{1}{c|}{\multirow{-2}{*}{\textbf{Video (Section~\ref{sec:video})}}}                                                                                                                                                                                                   & \multirow{-2}{*}{\textbf{Metrics}}                                          & \multicolumn{1}{c|}{\multirow{-2}{*}{\textbf{1}}} & \multicolumn{1}{c|}{\multirow{-2}{*}{\textbf{2}}} & \multicolumn{1}{c|}{\multirow{-2}{*}{\textbf{3}}} & \multicolumn{1}{c|}{\multirow{-2}{*}{\textbf{4}}} & \multicolumn{1}{c|}{\multirow{-2}{*}{\textbf{5}}} & \multicolumn{1}{c|}{\multirow{-2}{*}{-}}          & \multirow{-2}{*}{\textbf{Evaluation Dimensions}} & \multirow{-2}{*}{\textbf{Representative Works}}        \\ \hline
\rowcolor[HTML]{E1E0FF} 
\cellcolor[HTML]{E1E0FF}                                                                                                                                                                                                                                & CLIP-I~\cite{radford2021learning}                                                                      & \multicolumn{1}{c|}{\cellcolor[HTML]{E1E0FF}$\checkmark$}    & \multicolumn{1}{c|}{\cellcolor[HTML]{E1E0FF}$\checkmark$}    &                                                   & \multicolumn{1}{c|}{\cellcolor[HTML]{E1E0FF}$\checkmark$}    & \multicolumn{1}{l|}{\cellcolor[HTML]{E1E0FF}}     & \cellcolor[HTML]{E1E0FF}                          & Personalization                                  & PIA~\cite{zhang2024pia}, PoseCrafter~\cite{zhong2025posecrafter}, ID-Animator~\cite{he2024id}                          \\ \cline{2-7} \cline{9-10} 
\rowcolor[HTML]{E1E0FF} 
\cellcolor[HTML]{E1E0FF}                                                                                                                                                                                                                                & DINO-I~\cite{caron2021emerging,oquab2024dinov}                                                                      & \multicolumn{1}{c|}{\cellcolor[HTML]{E1E0FF}$\checkmark$}    & \multicolumn{1}{c|}{\cellcolor[HTML]{E1E0FF}$\checkmark$}    &                                                   & \multicolumn{1}{l|}{\cellcolor[HTML]{E1E0FF}}     & \multicolumn{1}{l|}{\cellcolor[HTML]{E1E0FF}}     & \cellcolor[HTML]{E1E0FF}                          & Personalization                                  & DisenStudio~\cite{chen2024disenstudio}, DreamVideo~\cite{wei2024dreamvideo}, Magic-Me~\cite{ma2024magic}                      \\ \cline{2-7} \cline{9-10} 
\rowcolor[HTML]{E1E0FF} 
\cellcolor[HTML]{E1E0FF}                                                                                                                                                                                                                                & SSIM~\cite{wang2004image}                                                                        &                                                   &                                                   & \multicolumn{1}{c|}{\cellcolor[HTML]{E1E0FF}$\checkmark$}    & \multicolumn{1}{c|}{\cellcolor[HTML]{E1E0FF}$\checkmark$}    & \multicolumn{1}{c|}{\cellcolor[HTML]{E1E0FF}$\checkmark$}    & \cellcolor[HTML]{E1E0FF}                          & Personalization                                  & AnimateAnyone~\cite{hu2024animate}, VITON-DiT~\cite{zheng2024viton}, ViViD~\cite{fang2024vivid}                        \\ \cline{2-7} \cline{9-10} 
\rowcolor[HTML]{E1E0FF} 
\cellcolor[HTML]{E1E0FF}                                                                                                                                                                                                                                & PSNR~\cite{hore2010image}                                                                        &                                                   &                                                   & \multicolumn{1}{c|}{\cellcolor[HTML]{E1E0FF}$\checkmark$}    & \multicolumn{1}{c|}{\cellcolor[HTML]{E1E0FF}$\checkmark$}    & \multicolumn{1}{c|}{\cellcolor[HTML]{E1E0FF}$\checkmark$}    & \cellcolor[HTML]{E1E0FF}                          & Personalization                                  & AnimateAnyone~\cite{hu2024animate}, \citet{yi2020audio}, \citet{zhua2023audio}                                \\ \cline{2-7} \cline{9-10} 
\rowcolor[HTML]{E1E0FF} 
\cellcolor[HTML]{E1E0FF}                                                                                                                                                                                                                                & LPIPS~\cite{zhang2018unreasonable}                                                                       &                                                   &                                                   & \multicolumn{1}{c|}{\cellcolor[HTML]{E1E0FF}$\checkmark$}    & \multicolumn{1}{c|}{\cellcolor[HTML]{E1E0FF}$\checkmark$}    & \multicolumn{1}{c|}{\cellcolor[HTML]{E1E0FF}$\checkmark$}    & \cellcolor[HTML]{E1E0FF}                          & Personalization                                  & AnimateAnyone~\cite{hu2024animate}, DiffTalk~\cite{shen2023difftalk}, DisCo~\cite{wang2023disco}                         \\ \cline{2-7} \cline{9-10} 
\rowcolor[HTML]{E1E0FF} 
\cellcolor[HTML]{E1E0FF}                                                                                                                                                                                                                                & VGG~\cite{johnson2016perceptual}                                                                         &                                                   &                                                   &                                                   & \multicolumn{1}{c|}{\cellcolor[HTML]{E1E0FF}$\checkmark$}    & \multicolumn{1}{l|}{\cellcolor[HTML]{E1E0FF}}     & \cellcolor[HTML]{E1E0FF}                          & Personalization                                  & DreamPose~\cite{karras2023dreampose}                                              \\ \cline{2-7} \cline{9-10} 
\rowcolor[HTML]{E1E0FF} 
\cellcolor[HTML]{E1E0FF}                                                                                                                                                                                                                                & L1 error                                                                    &                                                   &                                                   &                                                   & \multicolumn{1}{c|}{\cellcolor[HTML]{E1E0FF}$\checkmark$}    & \multicolumn{1}{l|}{\cellcolor[HTML]{E1E0FF}}     & \cellcolor[HTML]{E1E0FF}                          & Personalization                                  & DisCo~\cite{wang2023disco}, DreamPose~\cite{karras2023dreampose}, MagicAnimate~\cite{xu2024magicanimate}                         \\ \cline{2-7} \cline{9-10} 
\rowcolor[HTML]{E1E0FF} 
\cellcolor[HTML]{E1E0FF}                                                                                                                                                                                                                                & AED                                                                         &                                                   &                                                   &                                                   & \multicolumn{1}{c|}{\cellcolor[HTML]{E1E0FF}$\checkmark$}    & \multicolumn{1}{l|}{\cellcolor[HTML]{E1E0FF}}     & \cellcolor[HTML]{E1E0FF}                          & Personalization                                  & DisCo~\cite{wang2023disco}, DreamPose~\cite{karras2023dreampose}                                       \\ \cline{2-7} \cline{9-10} 
\rowcolor[HTML]{E1E0FF} 
\cellcolor[HTML]{E1E0FF}                                                                                                                                                                                                                                & Face similarity~\cite{deng2019arcface,huang2020curricularface,kim2022adaface}                                                             &                                                   & \multicolumn{1}{c|}{\cellcolor[HTML]{E1E0FF}$\checkmark$}    & \multicolumn{1}{c|}{\cellcolor[HTML]{E1E0FF}$\checkmark$}    & \multicolumn{1}{c|}{\cellcolor[HTML]{E1E0FF}$\checkmark$}    & \multicolumn{1}{l|}{\cellcolor[HTML]{E1E0FF}}     & \cellcolor[HTML]{E1E0FF}                          & Personalization                                  & ID-Animator~\cite{he2024id}, MagicPose~\cite{chang2023magicpose}, ConsisID~\cite{yuan2024identity}                       \\ \cline{2-7} \cline{9-10} 
\rowcolor[HTML]{E1E0FF} 
\cellcolor[HTML]{E1E0FF}                                                                                                                                                                                                                                & CLIP-T~\cite{radford2021learning}                                                                      & \multicolumn{1}{c|}{\cellcolor[HTML]{E1E0FF}$\checkmark$}    & \multicolumn{1}{c|}{\cellcolor[HTML]{E1E0FF}$\checkmark$}    &                                                   & \multicolumn{1}{c|}{\cellcolor[HTML]{E1E0FF}$\checkmark$}    & \multicolumn{1}{l|}{\cellcolor[HTML]{E1E0FF}}     & \cellcolor[HTML]{E1E0FF}                          & Instruction Alignment                            & PIA~\cite{zhang2024pia}, ConsisID~\cite{yuan2024identity}, PoseCrafter~\cite{zhong2025posecrafter}                             \\ \cline{2-7} \cline{9-10} 
\rowcolor[HTML]{E1E0FF} 
\cellcolor[HTML]{E1E0FF}                                                                                                                                                                                                                                & UMT score~\cite{liu2022umt}                                                                   & \multicolumn{1}{c|}{\cellcolor[HTML]{E1E0FF}$\checkmark$}    &                                                   &                                                   & \multicolumn{1}{l|}{\cellcolor[HTML]{E1E0FF}}     & \multicolumn{1}{l|}{\cellcolor[HTML]{E1E0FF}}     & \cellcolor[HTML]{E1E0FF}                          & Instruction Alignment                            & StyleMaster~\cite{ye2024stylemaster}                                            \\ \cline{2-7} \cline{9-10} 
\rowcolor[HTML]{E1E0FF} 
\cellcolor[HTML]{E1E0FF}                                                                                                                                                                                                                                & AKD~\cite{siarohin2021motion}                                                                         &                                                   &                                                   &                                                   & \multicolumn{1}{c|}{\cellcolor[HTML]{E1E0FF}$\checkmark$}    & \multicolumn{1}{l|}{\cellcolor[HTML]{E1E0FF}}     & \cellcolor[HTML]{E1E0FF}                          & Instruction Alignment                            & MagicAnimate~\cite{xu2024magicanimate}                                           \\ \cline{2-7} \cline{9-10} 
\rowcolor[HTML]{E1E0FF} 
\cellcolor[HTML]{E1E0FF}                                                                                                                                                                                                                                & MKR~\cite{siarohin2021motion}                                                                         &                                                   &                                                   &                                                   & \multicolumn{1}{c|}{\cellcolor[HTML]{E1E0FF}$\checkmark$}    & \multicolumn{1}{l|}{\cellcolor[HTML]{E1E0FF}}     & \cellcolor[HTML]{E1E0FF}                          & Instruction Alignment                            & MagicAnimate~\cite{xu2024magicanimate}                                           \\ \cline{2-7} \cline{9-10} 
\rowcolor[HTML]{E1E0FF} 
\cellcolor[HTML]{E1E0FF}                                                                                                                                                                                                                                & MSE-P                                                                       &                                                   &                                                   &                                                   & \multicolumn{1}{c|}{\cellcolor[HTML]{E1E0FF}$\checkmark$}    & \multicolumn{1}{l|}{\cellcolor[HTML]{E1E0FF}}     & \cellcolor[HTML]{E1E0FF}                          & Instruction Alignment                            & PoseCrafter~\cite{zhong2025posecrafter}                                            \\ \cline{2-7} \cline{9-10} 
\rowcolor[HTML]{E1E0FF} 
\cellcolor[HTML]{E1E0FF}                                                                                                                                                                                                                                & SyncNet score~\cite{chung2017out}                                                               &                                                   &                                                   & \multicolumn{1}{c|}{\cellcolor[HTML]{E1E0FF}$\checkmark$}    & \multicolumn{1}{l|}{\cellcolor[HTML]{E1E0FF}}     & \multicolumn{1}{l|}{\cellcolor[HTML]{E1E0FF}}     & \cellcolor[HTML]{E1E0FF}                          & Instruction Alignment                            & DreamTalk~\cite{ma2023dreamtalk}, EMO~\cite{tian2025emo}, MEMO~\cite{zheng2024memo}                         \\ \cline{2-7} \cline{9-10} 
\rowcolor[HTML]{E1E0FF} 
\cellcolor[HTML]{E1E0FF}                                                                                                                                                                                                                                & LMD~\cite{chen2018lip}                                                                         &                                                   &                                                   & \multicolumn{1}{c|}{\cellcolor[HTML]{E1E0FF}$\checkmark$}    & \multicolumn{1}{l|}{\cellcolor[HTML]{E1E0FF}}     & \multicolumn{1}{l|}{\cellcolor[HTML]{E1E0FF}}     & \cellcolor[HTML]{E1E0FF}                          & Instruction Alignment                            & DFA-NeRF~\cite{yao2022dfa}, DreamTalk~\cite{ma2023dreamtalk}, \citet{yi2020audio}                          \\ \cline{2-7} \cline{9-10} 
\rowcolor[HTML]{E1E0FF} 
\cellcolor[HTML]{E1E0FF}                                                                                                                                                                                                                                & LSE-C~\cite{lipsync}                                                                       &                                                   &                                                   & \multicolumn{1}{c|}{\cellcolor[HTML]{E1E0FF}$\checkmark$}    & \multicolumn{1}{l|}{\cellcolor[HTML]{E1E0FF}}     & \multicolumn{1}{l|}{\cellcolor[HTML]{E1E0FF}}     & \cellcolor[HTML]{E1E0FF}                          & Instruction Alignment                            & StyleLipSync~\cite{ki2023stylelipsync}, DiffTalker~\cite{qi2023difftalker}, \citet{choi2024text}                          \\ \cline{2-7} \cline{9-10} 
\rowcolor[HTML]{E1E0FF} 
\cellcolor[HTML]{E1E0FF}                                                                                                                                                                                                                                & LSE-D~\cite{lipsync}                                                                       &                                                   &                                                   & \multicolumn{1}{c|}{\cellcolor[HTML]{E1E0FF}$\checkmark$}    & \multicolumn{1}{l|}{\cellcolor[HTML]{E1E0FF}}     & \multicolumn{1}{l|}{\cellcolor[HTML]{E1E0FF}}     & \cellcolor[HTML]{E1E0FF}                          & Instruction Alignment                            & StyleLipSync~\cite{ki2023stylelipsync}, DiffTalker~\cite{qi2023difftalker}, \citet{choi2024text}                          \\ \cline{2-7} \cline{9-10} 
\rowcolor[HTML]{E1E0FF} 
\cellcolor[HTML]{E1E0FF}                                                                                                                                                                                                                                & PD~\cite{baldrati2023multimodal}                                                                          &                                                   &                                                   &                                                   & \multicolumn{1}{l|}{\cellcolor[HTML]{E1E0FF}}     & \multicolumn{1}{c|}{\cellcolor[HTML]{E1E0FF}$\checkmark$}    & \cellcolor[HTML]{E1E0FF}                          & Instruction Alignment                            & ACF~\cite{yang2024animated}                                                    \\ \cline{2-7} \cline{9-10} 
\rowcolor[HTML]{E1E0FF} 
\cellcolor[HTML]{E1E0FF}                                                                                                                                                                                                                                & FID~\cite{heusel2017gans}                                                                         &                                                   & \multicolumn{1}{c|}{\cellcolor[HTML]{E1E0FF}$\checkmark$}    & \multicolumn{1}{c|}{\cellcolor[HTML]{E1E0FF}$\checkmark$}    & \multicolumn{1}{c|}{\cellcolor[HTML]{E1E0FF}$\checkmark$}    & \multicolumn{1}{c|}{\cellcolor[HTML]{E1E0FF}$\checkmark$}    & \cellcolor[HTML]{E1E0FF}                          & Content Quality                                  & EMO~\cite{tian2025emo}, ConsisID~\cite{yuan2024identity}, DisCo~\cite{wang2023disco}                                   \\ \cline{2-7} \cline{9-10} 
\rowcolor[HTML]{E1E0FF} 
\cellcolor[HTML]{E1E0FF}                                                                                                                                                                                                                                & KID~\cite{binkowski2018demystifying}                                                                         &                                                   &                                                   &                                                   & \multicolumn{1}{l|}{\cellcolor[HTML]{E1E0FF}}     & \multicolumn{1}{c|}{\cellcolor[HTML]{E1E0FF}$\checkmark$}    & \cellcolor[HTML]{E1E0FF}                          & Content Quality                                  & WildVidFit~\cite{he2025wildvidfit}                                             \\ \cline{2-7} \cline{9-10} 
\rowcolor[HTML]{E1E0FF} 
\cellcolor[HTML]{E1E0FF}                                                                                                                                                                                                                                & ArtFID~\cite{wright2022artfid}                                                                      & \multicolumn{1}{c|}{\cellcolor[HTML]{E1E0FF}$\checkmark$}    &                                                   &                                                   & \multicolumn{1}{l|}{\cellcolor[HTML]{E1E0FF}}     & \multicolumn{1}{l|}{\cellcolor[HTML]{E1E0FF}}     & \cellcolor[HTML]{E1E0FF}                          & Content Quality                                  & StyleMaster~\cite{ye2024stylemaster}                                            \\ \cline{2-7} \cline{9-10} 
\rowcolor[HTML]{E1E0FF} 
\cellcolor[HTML]{E1E0FF}                                                                                                                                                                                                                                & VFID~\cite{vfid}                                                                        &                                                   &                                                   &                                                   & \multicolumn{1}{l|}{\cellcolor[HTML]{E1E0FF}}     & \multicolumn{1}{c|}{\cellcolor[HTML]{E1E0FF}$\checkmark$}    & \cellcolor[HTML]{E1E0FF}                          & Content Quality                                  & SwiftTry~\cite{nguyen2024swifttry}, VITON-DiT~\cite{zheng2024viton}, ViViD~\cite{fang2024vivid}                             \\ \cline{2-7} \cline{9-10} 
\rowcolor[HTML]{E1E0FF} 
\cellcolor[HTML]{E1E0FF}                                                                                                                                                                                                                                & FVD~\cite{unterthiner2018towards}                                                                         & \multicolumn{1}{c|}{\cellcolor[HTML]{E1E0FF}$\checkmark$}    & \multicolumn{1}{c|}{\cellcolor[HTML]{E1E0FF}$\checkmark$}    & \multicolumn{1}{c|}{\cellcolor[HTML]{E1E0FF}$\checkmark$}    & \multicolumn{1}{c|}{\cellcolor[HTML]{E1E0FF}$\checkmark$}    & \multicolumn{1}{l|}{\cellcolor[HTML]{E1E0FF}}     & \cellcolor[HTML]{E1E0FF}                          & Content Quality                                  & PersonalVideo~\cite{li2024personalvideo}, MotionBooth~\cite{wu2024motionbooth}, AnimateAnyone~\cite{hu2024animate}              \\ \cline{2-7} \cline{9-10} 
\rowcolor[HTML]{E1E0FF} 
\cellcolor[HTML]{E1E0FF}                                                                                                                                                                                                                                & FID-VID~\cite{balaji2019conditional}                                                                     &                                                   &                                                   &                                                   & \multicolumn{1}{c|}{\cellcolor[HTML]{E1E0FF}$\checkmark$}    & \multicolumn{1}{l|}{\cellcolor[HTML]{E1E0FF}}     & \cellcolor[HTML]{E1E0FF}                          & Content Quality                                  & DisCo~\cite{wang2023disco}, MagicAnimate~\cite{xu2024magicanimate}, MagicPose~\cite{chang2023magicpose}                         \\ \cline{2-7} \cline{9-10} 
\rowcolor[HTML]{E1E0FF} 
\cellcolor[HTML]{E1E0FF}                                                                                                                                                                                                                                & KVD~\cite{unterthiner2018towards}                                                                         & \multicolumn{1}{c|}{\cellcolor[HTML]{E1E0FF}$\checkmark$}    &                                                   &                                                   & \multicolumn{1}{l|}{\cellcolor[HTML]{E1E0FF}}     & \multicolumn{1}{l|}{\cellcolor[HTML]{E1E0FF}}     & \cellcolor[HTML]{E1E0FF}                          & Content Quality                                  & Animate-A-Story~\cite{he2023animate}                                        \\ \cline{2-7} \cline{9-10} 
\rowcolor[HTML]{E1E0FF} 
\cellcolor[HTML]{E1E0FF}                                                                                                                                                                                                                                & E-FID~\cite{tian2025emo}                                                                       &                                                   &                                                   & \multicolumn{1}{c|}{\cellcolor[HTML]{E1E0FF}$\checkmark$}    & \multicolumn{1}{l|}{\cellcolor[HTML]{E1E0FF}}     & \multicolumn{1}{l|}{\cellcolor[HTML]{E1E0FF}}     & \cellcolor[HTML]{E1E0FF}                          & Content Quality                                  & EMO~\cite{tian2025emo}, EmotiveTalk~\cite{wang2024emotivetalk}                                       \\ \cline{2-7} \cline{9-10} 
\rowcolor[HTML]{E1E0FF} 
\cellcolor[HTML]{E1E0FF}                                                                                                                                                                                                                                & NIQE~\cite{mittal2012making}                                                                        &                                                   &                                                   &                                                   & \multicolumn{1}{c|}{\cellcolor[HTML]{E1E0FF}$\checkmark$}    & \multicolumn{1}{l|}{\cellcolor[HTML]{E1E0FF}}     & \cellcolor[HTML]{E1E0FF}                          & Content Quality                                  & MagicFight~\cite{huang2024magicfight}                                             \\ \cline{2-7} \cline{9-10} 
\rowcolor[HTML]{E1E0FF} 
\cellcolor[HTML]{E1E0FF}                                                                                                                                                                                                                                & CPBD~\cite{narvekar2011no}                                                                        &                                                   &                                                   & \multicolumn{1}{c|}{\cellcolor[HTML]{E1E0FF}$\checkmark$}    & \multicolumn{1}{l|}{\cellcolor[HTML]{E1E0FF}}     & \multicolumn{1}{l|}{\cellcolor[HTML]{E1E0FF}}     & \cellcolor[HTML]{E1E0FF}                          & Content Quality                                  & DreamTalk~\cite{ma2023dreamtalk}                                              \\ \cline{2-7} \cline{9-10} 
\rowcolor[HTML]{E1E0FF} 
\cellcolor[HTML]{E1E0FF}                                                                                                                                                                                                                                & Temporal consistency~\cite{radford2021learning}                                                        & \multicolumn{1}{c|}{\cellcolor[HTML]{E1E0FF}$\checkmark$}    & \multicolumn{1}{c|}{\cellcolor[HTML]{E1E0FF}$\checkmark$}    &                                                   & \multicolumn{1}{l|}{\cellcolor[HTML]{E1E0FF}}     & \multicolumn{1}{l|}{\cellcolor[HTML]{E1E0FF}}     & \cellcolor[HTML]{E1E0FF}                          & Content Quality                                  & AnimateDiff~\cite{guo2024animatediff}, Magic-Me~\cite{ma2024magic}, DreamVideo~\cite{wei2024dreamvideo}                      \\ \cline{2-7} \cline{9-10} 
\rowcolor[HTML]{E1E0FF} 
\cellcolor[HTML]{E1E0FF}                                                                                                                                                                                                                                & Dynamic degree~\cite{huang2024vbench}                                                              & \multicolumn{1}{c|}{\cellcolor[HTML]{E1E0FF}$\checkmark$}    & \multicolumn{1}{c|}{\cellcolor[HTML]{E1E0FF}$\checkmark$}    &                                                   & \multicolumn{1}{l|}{\cellcolor[HTML]{E1E0FF}}     & \multicolumn{1}{l|}{\cellcolor[HTML]{E1E0FF}}     & \cellcolor[HTML]{E1E0FF}                          & Content Quality                                  & StyleMaster~\cite{ye2024stylemaster}, ID-Animator~\cite{he2024id}, PersonalVideo~\cite{li2024personalvideo}                \\ \cline{2-7} \cline{9-10} 
\rowcolor[HTML]{E1E0FF} 
\cellcolor[HTML]{E1E0FF}                                                                                                                                                                                                                                & Video IS~\cite{saito2020train}                                                                          & \multicolumn{1}{c|}{\cellcolor[HTML]{E1E0FF}$\checkmark$}    &                                                   &                                                   & \multicolumn{1}{l|}{\cellcolor[HTML]{E1E0FF}}     & \multicolumn{1}{c|}{\cellcolor[HTML]{E1E0FF}$\checkmark$}    & \cellcolor[HTML]{E1E0FF}                          & Content Quality                                  & MagDiff~\cite{zhao2025magdiff}                                 \\ \cline{2-7} \cline{9-10} 
\rowcolor[HTML]{E1E0FF} 
\cellcolor[HTML]{E1E0FF}                                                                                                                                                                                                                                & Flow error~\cite{shi2023videoflow}                                                                  & \multicolumn{1}{c|}{\cellcolor[HTML]{E1E0FF}$\checkmark$}    &                                                   &                                                   & \multicolumn{1}{l|}{\cellcolor[HTML]{E1E0FF}}     & \multicolumn{1}{l|}{\cellcolor[HTML]{E1E0FF}}     & \cellcolor[HTML]{E1E0FF}                          & Content Quality                                  & MotionBooth~\cite{wu2024motionbooth}                                            \\ \cline{2-7} \cline{9-10} 
\rowcolor[HTML]{E1E0FF} 
\cellcolor[HTML]{E1E0FF}                                                                                                                                                                                                                                & Stitch score                                                                & \multicolumn{1}{c|}{\cellcolor[HTML]{E1E0FF}$\checkmark$}    &                                                   &                                                   & \multicolumn{1}{l|}{\cellcolor[HTML]{E1E0FF}}     & \multicolumn{1}{l|}{\cellcolor[HTML]{E1E0FF}}     & \cellcolor[HTML]{E1E0FF}                          & Content Quality                                  & VideoDreamer~\cite{chen2023videodreamer}                                           \\ \cline{2-7} \cline{9-10} 
\rowcolor[HTML]{E1E0FF} 
\cellcolor[HTML]{E1E0FF}                                                                                                                                                                                                                                & Dover score~\cite{wu2023exploring}                                                                 &                                                   & \multicolumn{1}{c|}{\cellcolor[HTML]{E1E0FF}$\checkmark$}    &                                                   & \multicolumn{1}{l|}{\cellcolor[HTML]{E1E0FF}}     & \multicolumn{1}{l|}{\cellcolor[HTML]{E1E0FF}}     & \cellcolor[HTML]{E1E0FF}                          & Content Quality                                  & ID-Animator~\cite{he2024id}                                            \\ \cline{2-7} \cline{9-10} 
\rowcolor[HTML]{E1E0FF} 
\multirow[t]{-57}{*}{\cellcolor[HTML]{E1E0FF}\textbf{\begin{tabular}[c]{@{}l@{}}1. Subject-driven T2V generation\\ 2. ID-preserving T2V generation\\ 3. Talking head generation\\ 4. Pose-guided video generation\\ 5. Video virtual try-on\end{tabular}}} & Motion score~\cite{li2018vmaf}                                                                &                                                   & \multicolumn{1}{c|}{\cellcolor[HTML]{E1E0FF}$\checkmark$}    &                                                   & \multicolumn{1}{l|}{\cellcolor[HTML]{E1E0FF}}     & \multicolumn{1}{l|}{\cellcolor[HTML]{E1E0FF}}     & \multirow{-36}{*}{\cellcolor[HTML]{E1E0FF}}       & Content Quality                                  & ID-Animator~\cite{he2024id}                                            \\ \hline
\multicolumn{1}{c|}{}                                                                                                                                                                                                                                   &                                                                             & \multicolumn{1}{c|}{}                             & \multicolumn{1}{c|}{}                             & \multicolumn{1}{c|}{}                             & \multicolumn{1}{c|}{}                             & \multicolumn{1}{c|}{}                             & \multicolumn{1}{c|}{}                             &                                                  &                                                        \\
\multicolumn{1}{c|}{\multirow{-2}{*}{\textbf{3D (Section~\ref{sec:3D})}}}                                                                                                                                                                                                      & \multirow{-2}{*}{\textbf{Metrics}}                                          & \multicolumn{1}{c|}{\multirow{-2}{*}{\textbf{1}}} & \multicolumn{1}{c|}{\multirow{-2}{*}{\textbf{2}}} & \multicolumn{1}{c|}{\multirow{-2}{*}{\textbf{3}}} & \multicolumn{1}{c|}{\multirow{-2}{*}{\textbf{4}}} & \multicolumn{1}{c|}{\multirow{-2}{*}{-}}          & \multicolumn{1}{c|}{\multirow{-2}{*}{-}}          & \multirow{-2}{*}{\textbf{Evaluation Dimensions}} & \multirow{-2}{*}{\textbf{Representative Works}}        \\ \hline
\rowcolor[HTML]{EEFFE8} 
\cellcolor[HTML]{EEFFE8}                                                                                                                                                                                                                                & LPIPS~\cite{zhang2018unreasonable}                                                                       & \multicolumn{1}{c|}{\cellcolor[HTML]{EEFFE8}$\checkmark$}    & \multicolumn{1}{c|}{\cellcolor[HTML]{EEFFE8}$\checkmark$}    & \multicolumn{1}{c|}{\cellcolor[HTML]{EEFFE8}$\checkmark$}    & \multicolumn{1}{l|}{\cellcolor[HTML]{EEFFE8}}     & \multicolumn{2}{l|}{\cellcolor[HTML]{EEFFE8}}                                                         & Personalization                                  & Wonder3D~\cite{long2024wonder3d}, PuzzleAvatar~\cite{xiu2024puzzleavatar}, My3DGen~\cite{qi2023my3dgen} \\ \cline{2-6} \cline{9-10} 
\rowcolor[HTML]{EEFFE8} 
\cellcolor[HTML]{EEFFE8}                                                                                                                                                                                                                                & PSNR~\cite{hore2010image}                                                                        & \multicolumn{1}{c|}{\cellcolor[HTML]{EEFFE8}$\checkmark$}    & \multicolumn{1}{c|}{\cellcolor[HTML]{EEFFE8}$\checkmark$}    & \multicolumn{1}{c|}{\cellcolor[HTML]{EEFFE8}$\checkmark$}    & \multicolumn{1}{l|}{\cellcolor[HTML]{EEFFE8}}     & \multicolumn{2}{l|}{\cellcolor[HTML]{EEFFE8}}                                                         & Personalization                                  & Wonder3D~\cite{long2024wonder3d}, PuzzleAvatar~\cite{xiu2024puzzleavatar}, My3DGen~\cite{qi2023my3dgen}                        \\ \cline{2-6} \cline{9-10} 
\rowcolor[HTML]{EEFFE8} 
\cellcolor[HTML]{EEFFE8}                                                                                                                                                                                                                                & SSIM~\cite{wang2004image}                                                                        & \multicolumn{1}{c|}{\cellcolor[HTML]{EEFFE8}$\checkmark$}    & \multicolumn{1}{c|}{\cellcolor[HTML]{EEFFE8}$\checkmark$}    & \multicolumn{1}{c|}{\cellcolor[HTML]{EEFFE8}$\checkmark$}    & \multicolumn{1}{l|}{\cellcolor[HTML]{EEFFE8}}     & \multicolumn{2}{l|}{\cellcolor[HTML]{EEFFE8}}                                                         & Personalization                                  & Wonder3D~\cite{long2024wonder3d}, PuzzleAvatar~\cite{xiu2024puzzleavatar}, My3DGen~\cite{qi2023my3dgen} \\ \cline{2-6} \cline{9-10} 
\rowcolor[HTML]{EEFFE8} 
\cellcolor[HTML]{EEFFE8}                                                                                                                                                                                                                                & Chamfer Distances~\cite{butt1998optimum},                                                          & \multicolumn{1}{c|}{\cellcolor[HTML]{EEFFE8}$\checkmark$}    &                                                   &                                                   & \multicolumn{1}{l|}{\cellcolor[HTML]{EEFFE8}}     & \multicolumn{2}{l|}{\cellcolor[HTML]{EEFFE8}}                                                         & Personalization                                  & Wonder3D~\cite{long2024wonder3d}, PuzzleAvatar~\cite{xiu2024puzzleavatar}                                 \\ \cline{2-6} \cline{9-10} 
\rowcolor[HTML]{EEFFE8} 
\cellcolor[HTML]{EEFFE8}                                                                                                                                                                                                                                & CLIP~\cite{radford2021learning}                                                                        &                                                   &                                                   &                                                   & \multicolumn{1}{c|}{\cellcolor[HTML]{EEFFE8}$\checkmark$}    & \multicolumn{2}{l|}{\cellcolor[HTML]{EEFFE8}}                                                         & Personalization                                  & DreamVTON~\cite{xie2024dreamvton}                                              \\ \cline{2-6} \cline{9-10} 
\rowcolor[HTML]{EEFFE8} 
\cellcolor[HTML]{EEFFE8}                                                                                                                                                                                                                                & Volume IoU~\cite{zhou2019iou},                                                                 & \multicolumn{1}{c|}{\cellcolor[HTML]{EEFFE8}$\checkmark$}    &                                                   &                                                   & \multicolumn{1}{l|}{\cellcolor[HTML]{EEFFE8}}     & \multicolumn{2}{l|}{\cellcolor[HTML]{EEFFE8}}                                                         & Personalization                                  & Wonder3D~\cite{long2024wonder3d}                                               \\ \cline{2-6} \cline{9-10}
\rowcolor[HTML]{EEFFE8} 
\cellcolor[HTML]{EEFFE8}                                                                                                                                                                                                                                & Lip Vertex Error (LVE)~\cite{ma2024diffspeaker}                                                      &                                                   & \multicolumn{1}{c|}{\cellcolor[HTML]{EEFFE8}$\checkmark$}    &                                                   & \multicolumn{1}{l|}{\cellcolor[HTML]{EEFFE8}}     & \multicolumn{2}{l|}{\cellcolor[HTML]{EEFFE8}}                                                         & Personalization                                                 & DiffSpeaker~\cite{ma2024diffspeaker}                                            \\ \cline{2-6} \cline{9-10} 
\rowcolor[HTML]{EEFFE8} 
\cellcolor[HTML]{EEFFE8}                                                                                                                                                                                                                                & \begin{tabular}[c]{@{}l@{}}Facial Dynamics Devia-\\ tion (FDD)~\cite{ma2024diffspeaker}\end{tabular} &                                                   & \multicolumn{1}{c|}{\cellcolor[HTML]{EEFFE8}$\checkmark$}    &                                                   & \multicolumn{1}{l|}{\cellcolor[HTML]{EEFFE8}}     & \multicolumn{2}{l|}{\cellcolor[HTML]{EEFFE8}}                                                         & Personalization                                                 & DiffSpeaker~\cite{ma2024diffspeaker}, DiffsuionTalker~\cite{chen2023diffusiontalker}                           \\ \cline{2-6} \cline{9-10}
\rowcolor[HTML]{EEFFE8} 
\cellcolor[HTML]{EEFFE8}                                                                                                                                                                                                                                & FReID~\cite{huang2021few}                                                                       &                                                   &                                                   & \multicolumn{1}{c|}{\cellcolor[HTML]{EEFFE8}$\checkmark$}    & \multicolumn{1}{l|}{\cellcolor[HTML]{EEFFE8}}     & \multicolumn{2}{l|}{\multirow{-12}{*}{\cellcolor[HTML]{EEFFE8}}}                                      & Personalization                                                 & FewShotMotionTransfer~\cite{huang2021few}                                                \\ \cline{2-6} \cline{9-10}
\rowcolor[HTML]{EEFFE8} 
\cellcolor[HTML]{EEFFE8}                                                                                                                                                                                                                                & CLIP-T~\cite{radford2021learning}                                                                      & \multicolumn{1}{c|}{\cellcolor[HTML]{EEFFE8}$\checkmark$}    &                                                   &                                                   & \multicolumn{1}{l|}{\cellcolor[HTML]{EEFFE8}}     & \multicolumn{2}{l|}{\cellcolor[HTML]{EEFFE8}}                                                         & Instruction Alignment                            & MVDream~\cite{shi2023mvdream}, DreamBooth3D~\cite{raj2023dreambooth3d}, MakeYour3D~\cite{liu2025make}                      \\ \cline{2-6} \cline{9-10} 
\rowcolor[HTML]{EEFFE8} 
\cellcolor[HTML]{EEFFE8}                                                                                                                                                                                                                                & FID~\cite{heusel2017gans}                                                                         & \multicolumn{1}{c|}{\cellcolor[HTML]{EEFFE8}$\checkmark$}    &                                                   &                                                   & \multicolumn{1}{c|}{\cellcolor[HTML]{EEFFE8}$\checkmark$}    & \multicolumn{2}{l|}{\cellcolor[HTML]{EEFFE8}}                                                         & Content Quality                                  & MVDream~\cite{shi2023mvdream}, 3DAvatarGAN~\cite{abdal20233davatargan}, TextureDreamer~\cite{yeh2024texturedreamer}, DreamVTON~\cite{xie2024dreamvton}        \\ \cline{2-6} \cline{9-10}  
\rowcolor[HTML]{EEFFE8} 
\multirow[t]{-20}{*}{\cellcolor[HTML]{EEFFE8}\textbf{\begin{tabular}[c]{@{}l@{}}1. Image-to-3D generation\\ 2. 3D face generation\\ 3. 3D human pose generation\\ 4. 3D virtual try-on\end{tabular}}}                                                      & IS~\cite{salimans2016improved}                                                                          & \multicolumn{1}{c|}{\cellcolor[HTML]{EEFFE8}$\checkmark$}    &                                                   &                                                   & \multicolumn{1}{l|}{\cellcolor[HTML]{EEFFE8}}     & \multicolumn{2}{l|}{\cellcolor[HTML]{EEFFE8}}                                                         & Content Quality                                  & MVDream~\cite{shi2023mvdream}                                  \\ \hline
\multicolumn{1}{c|}{}                                                                                                                                                                                                                                   &                                                                             & \multicolumn{1}{c|}{}                             & \multicolumn{1}{c|}{}                             & \multicolumn{1}{c|}{}                             & \multicolumn{1}{c|}{}                             & \multicolumn{1}{c|}{}                             & \multicolumn{1}{c|}{}                             &                                                  &                                                        \\
\multicolumn{1}{c|}{\multirow{-2}{*}{\textbf{Audio (Section~\ref{sec:audio})}}}                                                                                                                                                                                                   & \multirow{-2}{*}{\textbf{Metrics}}                                          & \multicolumn{1}{c|}{\multirow{-2}{*}{\textbf{1}}} & \multicolumn{1}{c|}{\multirow{-2}{*}{\textbf{2}}} & \multicolumn{1}{c|}{\multirow{-2}{*}{\textbf{3}}} & \multicolumn{1}{c|}{\multirow{-2}{*}{-}}          & \multicolumn{1}{c|}{\multirow{-2}{*}{-}}          & \multicolumn{1}{c|}{\multirow{-2}{*}{-}}          & \multirow{-2}{*}{\textbf{Evaluation Dimensions}} & \multirow{-2}{*}{\textbf{Representative Works}}        \\ \hline
\rowcolor[HTML]{FFEAEE} 
\cellcolor[HTML]{FFEAEE}                                                                                                                                                                                                                                & CLAP~\cite{elizalde2023clap}                                                                        &                                                   &                                                   & \multicolumn{1}{c|}{\cellcolor[HTML]{FFEAEE}$\checkmark$}    & \multicolumn{3}{l|}{\cellcolor[HTML]{FFEAEE}}                                                                                                             & Personalization                                  & DB\&TI~\cite{plitsis2024investigating}                                                 \\ \cline{2-5} \cline{9-10} 
\rowcolor[HTML]{FFEAEE} 
\cellcolor[HTML]{FFEAEE}                                                                                                                                                                                                                                & Embedding Distance                                                          & \multicolumn{1}{c|}{\cellcolor[HTML]{FFEAEE}$\checkmark$}    & \multicolumn{1}{c|}{\cellcolor[HTML]{FFEAEE}$\checkmark$}    &                                                   & \multicolumn{3}{l|}{\cellcolor[HTML]{FFEAEE}}                                                                                                             & Personalization                                  & UMP~\cite{ma2022content}, FR-PSS~\cite{wang2022residual}                                            \\ \cline{2-5} \cline{9-10} 
\rowcolor[HTML]{FFEAEE} 
\cellcolor[HTML]{FFEAEE}                                                                                                                                                                                                                                & FAD~\cite{kilgour2018fr}                                                                         & \multicolumn{1}{c|}{\cellcolor[HTML]{FFEAEE}$\checkmark$}    & \multicolumn{1}{c|}{\cellcolor[HTML]{FFEAEE}$\checkmark$}    & \multicolumn{1}{c|}{\cellcolor[HTML]{FFEAEE}$\checkmark$}    & \multicolumn{3}{l|}{\cellcolor[HTML]{FFEAEE}}                                                                                                             & Personalization                                  & UIGAN~\cite{wang2024user}, DiffAVA~\cite{mo2023diffava}, DB\&TI~\cite{plitsis2024investigating}                                 \\ \cline{2-5} \cline{9-10} 
\rowcolor[HTML]{FFEAEE} 
\cellcolor[HTML]{FFEAEE}                                                                                                                                                                                                                                & IS~\cite{salimans2016improved}                                                                          &                                                   &                                                   & \multicolumn{1}{c|}{\cellcolor[HTML]{FFEAEE}$\checkmark$}    & \multicolumn{3}{l|}{\cellcolor[HTML]{FFEAEE}}                                                                                                             & Content Quality                                  & DiffAVA~\cite{mo2023diffava}                                                \\ \cline{2-5} \cline{9-10} 
\rowcolor[HTML]{FFEAEE} 
\multirow[t]{-5}{*}{\cellcolor[HTML]{FFEAEE}\textbf{\begin{tabular}[c]{@{}l@{}}1. Face-to-speech generation\\ 2. Music generation\\ 3. Text-to-audio generation\end{tabular}}}                                                                             & STOI, ESTOI, PESQ~\cite{sheng2023zero}                                                          & \multicolumn{1}{c|}{\cellcolor[HTML]{FFEAEE}$\checkmark$}    &                                                   &                                                   & \multicolumn{3}{l|}{\multirow{-5}{*}{\cellcolor[HTML]{FFEAEE}}}                                                                                           & Content Quality                                  & Lip2Speech~\cite{sheng2023zero}                                             \\ \hline
\multicolumn{1}{c|}{}                                                                                                                                                                                                                                   &                                                                             & \multicolumn{1}{c|}{}                             & \multicolumn{1}{c|}{}                             & \multicolumn{1}{c|}{}                             & \multicolumn{1}{c|}{}                             & \multicolumn{1}{c|}{}                             & \multicolumn{1}{c|}{}                             &                                                  &                                                        \\
\multicolumn{1}{c|}{\multirow{-2}{*}{\textbf{Cross-modal (Section~\ref{sec:cross-modal})}}}                                                                                                                                                                                             & \multirow{-2}{*}{\textbf{Metrics}}                                          & \multicolumn{1}{c|}{\multirow{-2}{*}{\textbf{1}}} & \multicolumn{1}{c|}{\multirow{-2}{*}{\textbf{2}}} & \multicolumn{1}{c|}{\multirow{-2}{*}{\textbf{3}}} & \multicolumn{1}{c|}{\multirow{-2}{*}{-}}          & \multicolumn{1}{c|}{\multirow{-2}{*}{-}}          & \multicolumn{1}{c|}{\multirow{-2}{*}{-}}          & \multirow{-2}{*}{\textbf{Evaluation Dimensions}} & \multirow{-2}{*}{\textbf{Representative Works}}        \\ \hline
\rowcolor[HTML]{FFF8EA} 
\cellcolor[HTML]{FFF8EA}                                                                                                                                                                                                                                & BLEU,Meteor                                                                 &                                                   & \multicolumn{1}{c|}{\cellcolor[HTML]{FFF8EA}$\checkmark$}    & \multicolumn{1}{c|}{\cellcolor[HTML]{FFF8EA}$\checkmark$}    & \multicolumn{3}{l|}{\cellcolor[HTML]{FFF8EA}}                                                                                                             & Overall                                          & PVCG~\cite{wu2024understanding}, METER~\cite{geng2022improving}, PV-LLM~\cite{lin2024personalized}                                      \\ \cline{2-5} \cline{9-10} 
\rowcolor[HTML]{FFF8EA} 
\cellcolor[HTML]{FFF8EA}                                                                                                                                                                                                                                & Recall, Precision, F1                                                       &                                                   &                                                   & \multicolumn{1}{c|}{\cellcolor[HTML]{FFF8EA}$\checkmark$}    & \multicolumn{3}{l|}{\cellcolor[HTML]{FFF8EA}}                                                                                                             & Overall                                          & MyVLM~\cite{alaluf2025myvlm}, Yo'LLaVA~\cite{nguyen2024yollava}                                        \\ \cline{2-5} \cline{9-10} 
\rowcolor[HTML]{FFF8EA} 
\multirow[t]{-4}{*}{\cellcolor[HTML]{FFF8EA}\textbf{\begin{tabular}[c]{@{}l@{}}1. Robotics\\ 2. Caption/Comment generation\\ 3. Multimodal dialogue systems\end{tabular}}}                                                                                          & success rate                                                                & \multicolumn{1}{c|}{\cellcolor[HTML]{FFF8EA}$\checkmark$}    &                                                   &                                                   & \multicolumn{3}{l|}{\multirow{-3}{*}{\cellcolor[HTML]{FFF8EA}}}                                                                                           & Overall                                          & VPL~\cite{poddar2024personalizing}, Promptable Behaviors~\cite{hwang2024promptable}                              \\ \hline
\end{tabular}
}
\end{table*}
}

\subsection{Personalized Text Generation}
\label{sec:text}

Personalized text generation aims to provide textual content tailored to user preferences and needs, involving tasks ranging from information seeking to user simulation.

\subsubsection{User Behaviors}
User interactions with the system typically occur over time \cite{c211214d6611436d84b328c6c5580913}, allowing it to learn implicit preferences and behavioral patterns to enhance personalization and encourage long-term engagement \cite{Shi2013TowardsUL}. 
This personalized context is valuable for the following personalized text-based tasks.

\paragraph{Information Seeking} A primary use case of personalized text generation is crafting responses to align with user preferences, enabling more engaging interactions. The system can leverage user feedback (e.g., thumbs up/down and selected best responses) to tailor its responses to user preferences. While personalization has been extensively studied in the context of information access and search \cite{se-pqa-search, personal-information-hansi, personalized-web-search-challenge, aol4ps}, which aims to select a response from predefined options, it remains relatively underexplored in generative scenarios. This is largely due to the lack of standardized metrics and benchmarks. However, recently, \citet{li2024personalizedlanguagemodelingpersonalized} explored the use of personalized feedback to train LLMs to generate more tailored summaries for users, as a form of information seeking. \citet{kumar2024compocommunitypreferenceslanguage} extends this approach by collecting preference feedback from a group of users as a community to optimize the model's ability to respond to their collective information needs.

\paragraph{Recommendation} While recommendation is not directly involved in content generation, it plays a crucial role in delivering personalized content, which has been explored across various scenarios~\cite{amazon-recom-1,movie-lens,goodreads-1,mind-new-recom}. 
The use of generative models in Recommendation Systems (RecSys) has been widely studied \cite{bao2023bistepgroundingparadigmlarge, wu2024surveylargelanguagemodels, yang2023palrpersonalizationawarellms,lin2025order}. Specifically, LLMs have been utilized either through prompting \cite{lyu-etal-2024-llm} or by training them directly to perform recommendation tasks as a form of text generation \cite{lin2024dataefficientfinetuningllmbasedrecommendation}. Beyond transformer-based generative models, newer approaches like diffusion models have also been explored for recommendation, highlighting the versatility of generative methods in this domain \cite{diffrec, 10.5555/3666122.3667176}.

Other work has also leveraged realistic user interaction to explore personalization for review generation~\cite{amazon-recom-2,li2020generate,sun2020dual,li2021personalized} and news headline generation~\cite{ao2021pens, cai2023generating, song2023general}. For example, \citet{ao2021pens} presents a personalized headline generation benchmark by collecting user's click history on Microsoft News.

\subsubsection{User Documents}

In some cases, users may not interact with the system frequently but can provide valuable information for personalization, such as user-created documents \cite{salemi-etal-2024-lamp}. 

\paragraph{Writing Assistant} Personalization plays a critical role in enhancing text-based writing assistants, enabling tailored text generation across diverse formats and styles. For this purpose, the LaMP benchmark \cite{salemi-etal-2024-lamp, 10.1145/3626772.3657783, salemi2024comparingretrievalaugmentationparameterefficientfinetuning, zhuang2024hydra} focuses on short-form text generation, such as creating headlines for news articles or subject lines for emails. In contrast, the LongLaMP benchmark \cite{kumar2024longlampbenchmarkpersonalizedlongform} targets longer-form tasks, such as writing product reviews from a user’s perspective based on a rating, generating a post from its summary, or completing an email for a user \cite{salemi2025reasoningenhancedselftraininglongformpersonalized,liu2024llms+,qiu2025measuring}.  
Additionally, the Personalized Linguistic Attributes Benchmark \cite{alhafni-etal-2024-personalized} is designed for text completion tasks, such as extending a post or review, leveraging user-written documents to extract stylistic features that guide LLMs in mimicking a user’s unique writing style.
In this domain, RAG is the dominant approach, retrieving relevant information from users' historically written documents to extract their writing preferences~\cite{mysore2024pearlpersonalizinglargelanguage, nicolicioiu2024panzapersonalizedtextwriting, li2023teachllmspersonalize}.

\subsubsection{User Profiles}
Generative models can infer user preferences from their profiles to guide personalized text generation.

\paragraph{Dialogue System} In recent years, chatbots and conversational systems have been a central focus of personalized text generation \cite{ijcai2017p521, 10.5555/1622467.1622479, kaiss:hal-04080920, info:doi/10.2196/15360}. The advent of advanced chat models like GPT-4 and Gemini have enabled more sophisticated and personalized interactions. By defining a persona or personality for these models based on user profiles, the system can tailor responses to individual preferences and needs \cite{10.1145/3469595.3469607, ijcai2019p721}. To support research in this domain, various dialogue datasets have been developed. For example, LiveChat \cite{livechat} is a large-scale dataset created from live streaming interactions, FoCus \cite{Jang2021CallFC} focuses on conversational information-seeking scenarios, and Pchatbot \cite{10.1145/3404835.3463239} compiles data from Weibo and judicial forums. Enhancing LLMs' ability to generate personalized responses involves approaches such as zero-shot prompting \cite{zhu-etal-2023-paed}, in-context learning \cite{xu-etal-2023-towards-zero}, and training on limited persona datasets \cite{song-etal-2021-bob, wang2024unimsragunifiedmultisourceretrievalaugmented} or large-scale datasets \cite{Zheng_Zhang_Huang_Mao_2020, chen-etal-2023-towards-robust}. Additionally, chain-of-thought reasoning has proven effective in improving alignment with user preferences in dialogue systems \cite{10.1007/978-981-97-9431-7_17}.

\paragraph{User Simulation}  

Previous research demonstrates that LLMs excel at performing roles or personas assigned to them~\cite{shanahan2023roleplaylargelanguagemodels, chen2024from}, enabling user simulation based on their profiles to extract preferences and further personalize the system for them~\cite{magee2024dramamachinesimulatingcharacter, opinionqa}. In this domain, \citet{character-llm} develops a test playground to interview trained agents and assess whether they effectively memorize their assigned characters and experiences. Additionally, \citet{rolellm} introduced a large-scale dataset for evaluating the role-playing abilities of LLMs, while \citet{Ng2024HowWC} presented a dataset specifically designed for assessing these capabilities within video game contexts.

\subsubsection{Evaluation Metrics}
Evaluating personalized text generation is challenging because only the target user can accurately determine whether the generated content aligns with their preferences and needs. One approach to evaluating personalized text generation is through human judgment, where individuals assess the quality and relevance of the generated content. Automatic evaluation of personalized text generation can be conducted using reference-based and reference-free approaches. In reference-based evaluation, it is assumed that a reference output is available for comparison. This comparison can be performed using term-matching metrics such as accuracy, ROUGE \cite{lin-2004-rouge}, BLEU \cite{10.3115/1073083.1073135}, and METEOR \cite{banerjee-lavie-2005-meteor} or semantic-matching methods using models like BERTScore \cite{Zhang*2020BERTScore:}, GEMBA \cite{kocmi2023largelanguagemodelsstateoftheart}, or G-Eval \cite{liu-etal-2023-g}. Recently, ExPerT \cite{salemi2025experteffectiveexplainableevaluation} was introduced for evaluating personalized text generation in reference-based settings. It segments both the generated text and the reference output into atomic facts and scores them based on content and writing style similarity. In reference-free evaluation, an LLM can assess the generated content directly, assigning scores based on various aspects, including how well the content aligns with the user profile or preferences \cite{wang2023automatedevaluationpersonalizedtext, wang-etal-2024-learning-personalized}, offering a more dynamic and personalized evaluation framework.

\subsection{Personalized Image Generation}
\label{sec:image}
Personalized image generation aims to synthesize images that reflect individual preferences and requirements. By incorporating various personalized contexts, existing studies have made significant strides in enhancing the capability of generative models to produce images tailored to specific needs, ranging from general-purpose generation to more specialized tasks.

\subsubsection{User Behaviors}
User interactions serve as a key source for inferring visual preferences, guiding the personalized image generation process. Based on user behaviors such as historical engagements and real-time feedback, existing methods have explored various approaches to enhance personalization.

\paragraph{General-purpose Image Generation} This task involves generating tailored images across various scenarios.
For instance, PMG~\cite{shen2024pmg}, I-AM-G~\cite{wang2024g}, Pigeon~\cite{xu2024personalized}, and DRC~\cite{xu2025drc} utilize historically interacted images of users to infer their visual tastes, enabling personalized generation across various scenarios, including stickers, movie posters, fashion designs, and news posters. In specific, PMG converts historical interacted images into textual descriptions, allowing LLMs to distill user preferences effectively; I-AM-G introduces an interest rewrite strategy to address preference ambiguity and leverages RAG to enrich semantic information; and Pigeon leverages MLLMs with specialized modules to manage noisy historical data and multimodal instructions, ensuring precise user modeling. Similarly, SGDM~\cite{xu2024sgdm} enhances personalization by introducing a style extraction module that captures user-specific style preferences to guide image generation. Personalized PR~\cite{chen2024tailored} proposes a personalized prompt-rewriting method, leveraging historical user query-image pairs to enhance text-to-image (T2I) personalization.

Beyond inferring user preferences from interaction history, some studies~\cite{von2023fabric,liu2024you,nabati2024personalized} have explored interactive, personalized image generation by incorporating real-time user feedback through multi-turn interactions, thereby progressively refining the generated outputs.

\paragraph{Fashion Design Generation} This task involves personalized fashion design with inferring personal style preferences from user behaviors. \citet{yu2019personalized} employs GANs to learn user preference vectors from interaction history, generating fashion images compatible with provided images. DiFashion~\cite{xu2024diffusion} adopts diffusion models to extract user preferences from interaction history for personalized outfit generation and recommendation.

\paragraph{E-commerce Product Image Generation} This task aims to create customized, eye-catching visuals for e-commerce products to attract target consumers. Based on user behaviors, AdBooster~\cite{shilova2023adbooster} utilizes the Stable Diffusion outpainting model, conditioning on individual user interest to generate appealing product images. \citet{vashishtha2024chaining} leverages LLMs to generate text prompts for diffusion models to craft engaging banners. \citet{czapp2024dynamic} employs a contextual bandit algorithm to select prompts from a pool, generating personalized product backgrounds.

\subsubsection{User Profiles}
Some studies utilize users' demographic attributes to infer preferences or categorize them into groups for personalized image generation.

\paragraph{Fashion Design Generation} Based on user attributes (\eg age, gender, interests in characters), LVA-COG~\cite{forouzandehmehr2023character} utilizes LLMs to extract user preferences to guide fashion design generation. 
In contrast, PerFusion~\cite{lin2025sell} builds a reward model based on user profiles to estimate user preferences, which in turn guides group-level preference optimization of DMs for personalized fashion generation.

\paragraph{E-commerce Product Image Generation} By categorizing users into distinct groups based on their attributes, CG4CTR~\cite{yang2024new} proposes a self-cyclic generation pipeline to produce tailored product images for each user group.

\subsubsection{Personalized Subjects}
This is a primary focus of the computer vision community, which aims to capture the subject representation from a limited set of subject images and follow users' instructions for subject-driven text-to-image (T2I) generation. For instance, given a few images of a user's pet, the model generates new images featuring the pet in different contexts or environments while preserving its identity.

\paragraph{Subject-driven T2I Generation}
Existing research in this area can be broadly categorized into two branches: 
\begin{itemize}[leftmargin=*,itemsep=2pt,topsep=2pt,parsep=0pt]
    \item  Optimization-based methods
introduce a learnable unique identifier in the embedding space to encapsulate the semantics and visual details of each subject. Specifically, Textual Inversion~\cite{gal2023an} optimizes a pseudo-word identifier, denoted as $S^*$, to encode a personalized representation that guides T2I generation in diffusion models.
DreamBooth~\cite{ruiz2023dreambooth} combines a unique identifier with a subject class name (\eg ``A $S^*$ cat'') to leverage the class-specific prior knowledge embedded in the model, leveraging the model’s prior knowledge of the class. Building upon these pioneering works, recent efforts have aimed to improve efficiency, subject fidelity, and instruction alignment, addressing both single-subject generation~\cite{voynov2023p+,alaluf2023neural,tewel2023key,pang2024attndreambooth,cai2024decoupled,hong2025comfusion} and multi-subject generation~\cite{avrahami2023break,kumari2023multi,gu2024mix,yao2024concept,zhang2024compositional}.

    \item Encoder-based methods
utilize a pre-trained image encoder to extract subject-specific features, which are then incorporated into text prompts or directly injected into the generator through dedicated cross-attention mechanisms or adapters.
For instance, ELITE~\cite{wei2023elite}, IP-Adapter~\cite{ye2023ip}, 
Subject-Diffusion~\cite{ma2024subject} and SSR-Encoder~\cite{zhang2024ssr} employ CLIP~\cite{radford2021learning} as the feature extractor, each adopting unique encoding and injection strategies to seamlessly incorporate subject features into the image generation process. MoMA~\cite{song2025moma} takes advantage of a pre-trained MLLM, LLaVA~\cite{liu2023visual}, to extract subject features and refine them based on the target prompt, producing contextualized image features that are injected into the generator via cross-attention layers. In addition, encoder-based methods have been extended to support multi-subject generation~\cite{patel2024eclipse,li-etal-2024-unimo,zhu2024multibooth,wang2024ms}.
\end{itemize}
Furthermore, other research has explored diverse techniques for personalized subject-driven generation. These include prompt engineering\cite{he2024automated}, which formulates structured prompts to guide generation; instruction tuning\cite{zhou2024customization,hu2024instruct}, which fine-tunes models on personalized instructions for improved alignment; and reinforcement learning~\cite{chen2023subjectdriven,chae2023instructbooth,huang2024patchdpo,wei2025powerful}, which optimizes models through user feedback to refine subject representation.

Moreover, beyond capturing user preferences for concrete subjects like objects, some studies have focused on more abstract concepts, such as specified relations or styles, to guide personalized T2I generation~\cite{huang2023reversion, sohn2023styledrop, wang2023rethinking, liu2024ada, park2024text}. 

\subsubsection{Personal Face/Body}
Personal face and body images have become popular for personalized image generation, due to their high relevance to individual identity. By leveraging these images, generative models can create highly tailored and realistic images that reflect users' distinct identities (IDs) while adhering to user-specific requirements. Existing studies primarily focus on face generation and virtual try-on applications.

\paragraph{Face Generation}
Generative models utilize personal face images to create high-fidelity portraits or avatars that preserve individual face IDs while adhering to users' multimodal instructions, such as modifying expressions, actions, backgrounds, and styles.
Early GAN-based work mainly encodes face images into the latent space of StyleGAN~\cite{stylegan} for face manipulation~\cite{xia2021tedigan,patashnik2021styleclip,deltaedit,baykal2023clip}. 
To achieve more precise identity preservation and more flexible control, DM-based methods usually integrate a separate image encoder to convert face images into ID representations, which are then combined with user instructions to guide the face generation process. 
For instance, FastComposer~\cite{xiao2023fastcomposer}, Face0~\cite{valevski2023face0}, PhotoMaker~\cite{li2024photomaker}, Infinite-ID~\cite{wu2025infinite}, MasterWeaver~\cite{wei2025masterweaver}, and AddMe~\cite{yue2025addme} utilize a pre-trained CLIP image encoder or face recognition model to extract ID embeddings for identity preservation. In contrast, SeFi-IDE~\cite{li2024sefi} directly optimizes one ID representation as multiple per-stage tokens to enhance semantic control. 
Except for ID representations, some studies have explored additional features or conditions to enhance style control~\cite{yan2023facestudio}, spatial control~\cite{wang2024instantid,he2024imagine,he2024uniportrait,jiang2025groupdiff}, especially specific human-object interaction ~\cite{guo2024pulid, hu2025personahoi}, and scene affordance~\cite{kulal2023putting,parihar2024text2place}. 

However, the rapid development of personalized face generation techniques has raised concerns about potential misuse and privacy risks. Recent research has investigated unlearning-related methods~\cite{wu2025unlearning, 10.1145/3658644.3670398}, adversarial attack-based methods~\cite{van2023anti,xiao2023my,wan2024promptagnostic,onikubo2024high,liu2024rethinking} and watermark-based methods~\cite{liu2024countering} to protect user privacy.

\paragraph{Virtual Try-on} This task aims to synthesize a photorealistic image of a dress person by combining their body and face images with specified garments. Early GAN-based works~\cite{wang2018toward,dong2019towards,men2020controllable,choi2021viton,lee2022high} mainly follow a two-stage process. Initially, a dedicated warping module is employed to align garment images with the person's body shape. Subsequently, the reshaped garment is seamlessly blended with the person's image to generate the final try-on result. With the great success of DMs in various tasks, recent research has deployed their applications in virtual try-ons~\cite{zhang2024mmtryon,ning2024picture,kim2024stableviton,wan2025improving}. For example, LaDI-VTON~\cite{morelli2023ladi} and DCI-VTON~\cite{gou2023taming} explicitly warp clothes to match the person's body and then utilize DMs for blending. In contrast, TryOnDiffusion~\cite{zhu2023tryondiffusion} proposes a Parallel-UNet architecture, which performs implicit warping and blending in a unified process. Similarly, several subsequent studies~\cite{choi2025improving,xu2024ootdiffusion,sun2024outfitanyone,shen2024imagdressing} utilize parallel UNets for garment feature extraction and enhance blending via self-attention and cross-attention mechanisms. 

\subsubsection{Evaluation Metrics}
To assess the alignment between generated images and personalized contexts, as well as adherence to users' multimodal instructions, most studies rely on similarity metrics like Learned Perceptual Image Patch Similarity (LPIPS)~\cite{zhang2018unreasonable} and Structural Similarity Index Measure (SSIM)~\cite{wang2004image}. 
Additionally, pre-trained models such as CLIP~\cite{radford2021learning}, DINO~\cite{oquab2024dinov}, and various face recognition models~\cite{schroff2015facenet,deng2019arcface} are often used to extract image features for computing cosine similarity, enabling a more contextual evaluation of personalization and instruction alignment. Moreover, Stellar~\cite{achlioptas2023stellar} introduces specialized metrics designed for subject-driven image generation and human generation.

To quantify the quality and coherence of generated images, conventional metrics such as Fr\'{e}chet Inception Distance (FID)~\cite{heusel2017gans} and Kernel Inception Distance (KID)~\cite{binkowski2018demystifying} are also commonly employed.
Beyond quantitative evaluation, most studies present case studies and conduct human evaluations to assess the personalization and instruction alignment of the generated images.
In e-commerce scenarios, product image generation often incorporates online tests to evaluate model performance in real-world scenarios.

\subsection{Personalized Video Generation}
\label{sec:video}

Personalized video generation aims to produce tailored video content that reflects individual preferences, traits, and specific needs.

\subsubsection{Personalized Subjects} 
In some cases, users may provide one or several images of a personalized subject, such as an object or concept, along with a specified text prompt, requiring generative models to perform subject-driven text-to-video (T2V) generation. This task is conceptually similar to subject-driven T2I generation.

\paragraph{Subject-driven T2V Generation}
Given the great success of various personalized models in subject-driven T2I generation, methods such as AnimateDiff~\cite{guo2024animatediff}, PIA~\cite{zhang2024pia}, and Still-Moving~\cite{chefer2024still} adapt these models for T2V generation by incorporating motion and temporal dynamics through additional modules. MagDiff~\cite{zhao2025magdiff} enhances subject-driven video generation with three types of alignments. VideoBooth~\cite{jiang2024videobooth} proposes a coarse-to-fine manner to encode subject images into the generator. CustomCrafter~\cite{wu2024customcrafter} integrates a plug-and-play module with a dynamic weighted video sampling strategy to maintain motion generation and conceptual combination abilities during subject learning. Other studies introduce additional conditions, such as motion control~\cite{wu2024motionbooth,wei2024dreamvideo,wei2024dreamvideo2} and depth control~\cite{he2023animate}, enabling more flexible subject customization. Besides, some studies have explored multi-subject T2V generation~\cite{chen2023videodreamer,chen2024disenstudio,wang2024customvideo}. Beyond these, methods such as StyleCrafter\cite{liu2024stylecrafter} and StyleMaster~\cite{ye2024stylemaster} integrate specified style images as subjects, allowing for stylized T2V generation that tailors the video aesthetic to the desired style.

\subsubsection{Personal Face/Body}
Similarly, users may provide one or more personal face and body images, enabling generative models to synthesize personalized videos that preserve their identities while following multimodal instructions. These tasks include ID-preserving T2V generation, talking head generation, pose-guided video generation, and video virtual try-on.

\paragraph{ID-preserving T2V Generation} This task focuses on creating personalized videos that align with personal face IDs and specified text prompts. For example, Magic-Me~\cite{ma2024magic} builds upon Textual Inversion~\cite{gal2023an} to learn ID-specific representations to guide T2V generation, requiring separate training for each ID. ID-Animator~\cite{he2024id} employs a face adapter to encode ID-related information and incorporates it into the generator via cross-attention. ConsisID~\cite{yuan2024identity} decomposes facial information into frequency-aware features, which are integrated into Diffusion Transformers (DiT) for video generation. PersonalVideo~\cite{li2024personalvideo} applies direct supervision on T2V-generated videos, aligning model tuning with the inference process.

\paragraph{Talking Head Generation}
This task aims to synthesize lip-synchronized talking videos, typically driven by personal face images and corresponding audio clips. Recent research has explored diverse approaches, such as Neural Radiance Fields (NeRFs)~\cite{yao2022dfa,li2023efficient} and different backbone networks, including GANs~\cite{yi2020audio,ki2023stylelipsync,guan2023stylesync} and DMs~\cite{zhang2023dream,zhua2023audio,shen2023difftalk,liu2024multimodal,wei2024aniportrait,tian2025emo,tan2025edtalk,wang2024emotivetalk,zheng2024memo}. Beyond audio-driven generation, some studies have also investigated video-driven~\cite{zhang2023metaportrait} and text-driven~\cite{choi2024text} methods for talking head generation.

\paragraph{Pose-guided Video Generation}
Recent studies have explored adapting personal face and body images to match specific pose sequences through various condition mechanisms for video generation~\cite{wang2023disco,chang2023magicpose,karras2023dreampose,xu2024magicanimate,hu2024animate,zhong2025posecrafter}. Beyond single-person scenarios, MagicFight~\cite{huang2024magicfight} tackles the complexities of two-person martial arts combat video generation, addressing challenges such as identity confusion and action mismatches. 

\paragraph{Video Virtual Try-on}
This task seeks to seamlessly transfer a specified garment onto a person in a source video while preserving their motion and identity. Early GAN-based work~\cite{dong2019fw,zhong2021mv,jiang2022clothformer} primarily follows a two-stage workflow, warping the specified garment and blending it with the target person by a GAN generator. Recent studies utilize DMs for video try-ons, incorporating specialized modules for garment and pose encoding, along with dedicated condition mechanisms, such as Tunnel Try-on~\cite{xu2024tunnel}, ACF~\cite{yang2024animated}, GPD-VVTO~\cite{wang2024gpd}, VITON-DiT~\cite{zheng2024viton}, ViViD~\cite{fang2024vivid}, WildVidFit~\cite{he2025wildvidfit}, and SwiftTry~\cite{nguyen2024swifttry}.

\subsubsection{Evaluation Metrics}
To assess personalization and instruction alignment, similar to personalized image generation, existing studies commonly rely on similarity metrics such as LPIPS, SSIM, and Peak Signal-to-Noise Ratio (PSNR)~\cite{hore2010image}. Additionally, pre-trained image encoders like CLIP and DINO are frequently used to extract frame-level features and compute cosine similarity for quantitative evaluation.
There are some task-specific metrics, such as SyncNet score~\cite{casale2018gaussian}, which evaluates audio-visual synchronization quality for audio-driven talking head generation, and face similarity metrics for ID-preserving human generation, which are based on backbone face recognition models like ArcFace~\cite{deng2019arcface} and CurricularFace~\cite{huang2020curricularface}.

For overall video quality assessment, standard metrics include frame-level evaluations like FID and KID, as well as video-level metrics such as VFID~\cite{vfid}, FID-VID~\cite{balaji2019conditional}, FVD~\cite{unterthiner2018towards}, and KVD~\cite{unterthiner2018towards}. Additionally, some studies leverage CLIP-based cosine similarity between consecutive video frames to assess temporal consistency.

Beyond quantitative metrics, many studies complement their evaluations with qualitative case studies and human assessments to better capture personalization, instruction alignment, and overall video quality.

\subsection{Personalized 3D Generation}
\label{sec:3D}
Personalized 3D generation involves transforming users' personalized visual or textual contexts (e.g., body shapes, facial features, images, and prompts) into 3D assets. 

\subsubsection{Personalized Subjects}
The most common paradigm for personalized 3D generation involves users providing some image-based personalized subjects, and then generating the corresponding 3D assets.

\paragraph{Image-to-3D Generation}
Personalized image-to-3D generation focuses on creating 3D assets that accurately capture the geometry and appearance of given personalized subjects.  
3DAvatarGAN~\cite{abdal20233davatargan} introduces a cross-domain adaptation framework that aligns features from 2D-GANs with those of 3D-GANs. PuzzleAvatar~\cite{xiu2024puzzleavatar} utilizes an enhanced Score Distillation Sampling (SDS) technique to optimize the geometry and texture of 3D portraits. TextureDreamer~\cite{yeh2024texturedreamer} integrates geometric information using ControlNet, proposing a Geometry-Aware Personalized Score Distillation (PGSD) approach.

Some methods further ensure alignment with textual prompts during the 3D generation process. MVDream~\cite{shi2023mvdream} employs a multi-view diffusion model to generate consistent multi-view images based on text prompts. DreamBooth3D~\cite{raj2023dreambooth3d} combines DreamFusion and DreamBooth models within a three-stage optimization framework to enhance detail preservation and consistency through multi-view pseudo-data generation. Wonder3D~\cite{long2024wonder3d} introduces a cross-domain diffusion model leveraging multi-view cross-attention to produce detailed normal and color maps. DreamFont3D~\cite{li2024dreamfont3d} utilizes NeRF as the 3D representation, optimizing font geometry and texture with multi-view mask constraints and progressive weight adjustments. Make-your-3D~\cite{liu2025make} implements a joint optimization framework that combines identity-aware and subject-prior optimizations, aligning a 2D personalization model with a multi-view diffusion model for accurate 3D generation.

\subsubsection{Personal Face/Body}
In some cases, users may provide personal face and body inputs in the form of images or monocular videos, aiming to generate identity-preserving 3D assets.

\paragraph{3D Face Generation}
For 3D face generation, \cite{zhang20213d} introduces PoseGAN, a module designed to generate dynamic head poses. \cite{gao2022reconstructing} proposes a linear blend model based on multi-level voxel fields, representing expressions as neural radiance field bases. My3DGen~\cite{qi2023my3dgen} adapts the pre-trained EG3D model using Low-Rank Adaptation (LoRA) for parameter-efficient training on a limited set of personalized images. DiffusionTalker~\cite{chen2023diffusiontalker} employs contrastive learning to map speech features to personalized speaker identity. \cite{wang2024uvmap} integrates a face fusion module into a fine-tuned text-to-image diffusion model for identity-driven customization.
\cite{ko2024talk3d} utilizes VIVE3D to fine-tune the EG3D generator by inverting key frames from monocular videos. \cite{song2024talkingstyle} applies Cross-Modal Aggregation to blend style and facial motion features, ensuring alignment between generated facial expressions, speech, and styles. DiffSpeaker~\cite{ma2024diffspeaker} proposes a diffusion model-based Transformer architecture to enhance speech-to-facial-animation mapping. 
Zero-1-to-A~\cite{zhou2025zero} leverages pretrained video DMs to generate 4D avatars from a single face image, capturing both 3D spatial and temporal dynamics. In addition, FaceGPT~\cite{wang2024facegpt} utilizes MLLMs for personalized 3D face understanding and generation from both user-provided face images and textual descriptions.

\paragraph{3D Human Pose generation}
For 3D human pose generation, \cite{huang2021few} combines source image shape information with 2D key points to generate a personalized UV map. PGG~\cite{hu2023personalized} introduces a geometry-aware graph constructed from intermediate human mesh predictions, enabling personalized and dynamic pose generation. 3DHM~\cite{li2024synthesizing} and DreamWaltz~\cite{huang2024dreamwaltz} enable animating people from a single image or textual prompts.

\paragraph{3D Virtual Try-on}
3D Virtual Try-on enables the creation of high-quality, customized 3D models from minimal inputs, such as user images, clothing images, and textual prompts. \cite{chu20173d} addresses the precision requirements of personalized facial modeling for applications like eyeglass frame design, utilizing parametric modeling techniques for 3D face creation. Pergamo~\cite{casado2022pergamo} addresses the challenge of reconstructing 3D clothing from 2D images, employing semantic segmentation, normal prediction, and a parameterized clothing model to optimize coarse geometry and fine details through differentiable rendering. DreamVTON~\cite{xie2024dreamvton} incorporates Multi-Concept LoRA and Normal Style LoRA into Stable Diffusion, enabling the generation of pose-consistent, detail-rich clothing models.

\subsection{Evaluation Metrics}
To quantify personalization and instruction alignment in 3D generation, existing studies commonly use similarity metrics such as LPIPS, SSIM, PSNR, and CLIP scores, similar to those in image and video generation. Additionally, some task-specific scores can be evaluated through pre-trained models, such as facial attribute classifiers ~\cite{abdal20233davatargan, qi2023my3dgen}.

For 3D geometric quality, commonly used metrics include Chamfer Distance (CD)~\cite{gao2022reconstructing, xiu2024puzzleavatar} and Point-to-Surface Distance (P2S)~\cite{xiu2024puzzleavatar} for shape fidelity, as well as Normal Consistency and Volume IoU for surface detail and volumetric overlap~\cite{xie2024dreamvton}. Task-specific metrics such as Mean Per Joint Position Error (MPJPE) and Mean Per Vertex Error (MPVE) are frequently used in the human body and pose estimation tasks~\cite{hu2023personalized}.

Beyond objective metrics, qualitative assessments like user studies are often conducted to evaluate subjective aspects of 3D generation~\cite{zhang20213d, qi2023my3dgen, xie2024dreamvton, huang2021few}, including realism, texture photorealism, and shape-texture consistency.

\subsection{Personalized Audio Generation}
\label{sec:audio}
Personalized audio generation extracts users' auditory preferences to create tailored audio content, such as music and speech.

\subsubsection{User behaviors}
User behaviors on music, such as listening history and ratings, are important clues for inferring user preferences for personalization.

\paragraph{Music Generation}
Methods like UMP~\cite{ma2022content} and UP-Transformer~\cite{hu2022can} infer user preferences by analyzing listening histories and ratings. UIGAN~\cite{wang2024user} adopts an interactive approach to collect user feedback, progressively refining the generated music. \citet{ma2024symbolic} constructs a music community graph, where nodes represent users and songs, and edges capture relationships such as likes, subscriptions, and user interactions. Furthermore, \citet{plitsis2024investigating} adapts image-domain personalization techniques like Textual Inversion~\cite{gal2023an} and DreamBooth~\cite{ruiz2023dreambooth} to audio tasks, enhancing user-centric music generation.

\subsubsection{Personalized Subjects}
In some cases, users provide audio clips and aim to manipulate them using textual prompts.
\paragraph{Text-to-Audio Generation}
Personalized text-to-audio generation explores techniques for synthesizing tailored audio by aligning user preferences, textual descriptions, and contextual inputs.
YourTTS~\cite{casanova2022yourtts} focuses on zero-shot multi-speaker and multilingual training, allowing fast adaptation to unseen voices. Recent advancements such as Voicebox~\cite{le2023voicebox}, StyleTTS 2~\cite{li2023styletts}, Make-An-Audio~\cite{huang2023make}, DiffAVA~\cite{mo2023diffava}, and TAS~\cite{li2024tas}, utilizes DMs for personalized text-to-audio generation.
For instance, DiffAVA utilizes Contrastive Language-Audio Pretraining to align features across audio and visual-text modalities in the spatiotemporal domain.  TAS introduces a Semantic-Aware Fusion module to capture text-aware audio features, establishing nuanced perceptual relationships between text and audio inputs for more contextually aligned audio outputs.
Additionally, VALL-E X~\cite{zhang2023speak} introduces a cross-lingual neural codec language model that supports personalized speech synthesis in a different language for monolingual speakers. Style-Talker~\cite{li2024style} builds a spoken dialogue system on a pretrained audio LLM, integrating user input audio, transcribed chat history, and speech styles to generate personalized audio responses.

\subsubsection{Personal Face/Body}
Users may provide their facial images or videos, enabling generative models to extract speaker-specific attributes for customized speech generation.

\paragraph{Face-to-Speech Generation}
VioceMe~\cite{van2022voiceme} employs SpeakerNet to derive speaker embeddings and incorporates Global Style Tokens for modeling speech styles. FR-PSS~\cite{wang2022residual} enhances the extraction of speech features from facial characteristics using a residual guidance strategy, which integrates prior speech information for improved efficiency. Lip2Speech~\cite{sheng2023zero} leverages a Variational Autoencoder framework to disentangle speaker identity from linguistic content in videos, achieving fine-grained control over personalized speech synthesis via speaker embeddings.

\subsubsection{Evaluation Metrics}
To quantify personalization and stylistic alignment in tasks like music transfer and text-to-audio generation, existing studies commonly use similarity metrics such as CLAP scores, Pattern Similarity (PS), and Embedding Distance are commonly applied to assess how well the generated content aligns with target musical styles or speaker characteristics~\cite{sheng2023zero, plitsis2024investigating}. 

Audio quality assessment often involves perceptual metrics like Short-Time Objective Intelligibility (STOI) and Extended STOI (ESTOI) for speech intelligibility, Perceptual Evaluation of Speech Quality (PESQ) for clarity, and Fréchet Audio Distance (FAD) for measuring realism and diversity. For music generation tasks, metrics such as Precision, Recall, Density, and Coverage (P\&R\&D\&C) are frequently employed to capture both creativity and output diversity~\cite{wang2024user, mo2023diffava, hu2022can}.

Subjective evaluations remain crucial, particularly for tasks where user experience and personalization are central. User studies are often conducted to measure factors such as perceived musicality, naturalness, emotional impact, and how closely the generated content reflects user preferences~\cite{ma2022content, dai2022personalised}. 

\subsection{Personalized Cross-modal Generation}
\label{sec:cross-modal}
Personalized cross-modal generation primarily aims to produce personalized textual responses (e.g., captions, answers, or robot actions) based on multimodal personalized contexts (e.g., images, videos, historical robot trajectories). 

\begin{figure*}[t]
% \vspace{-0.2cm}
\setlength{\abovecaptionskip}{0.05cm}
\setlength{\belowcaptionskip}{-0.45cm}
\centering
\includegraphics[scale=0.57]{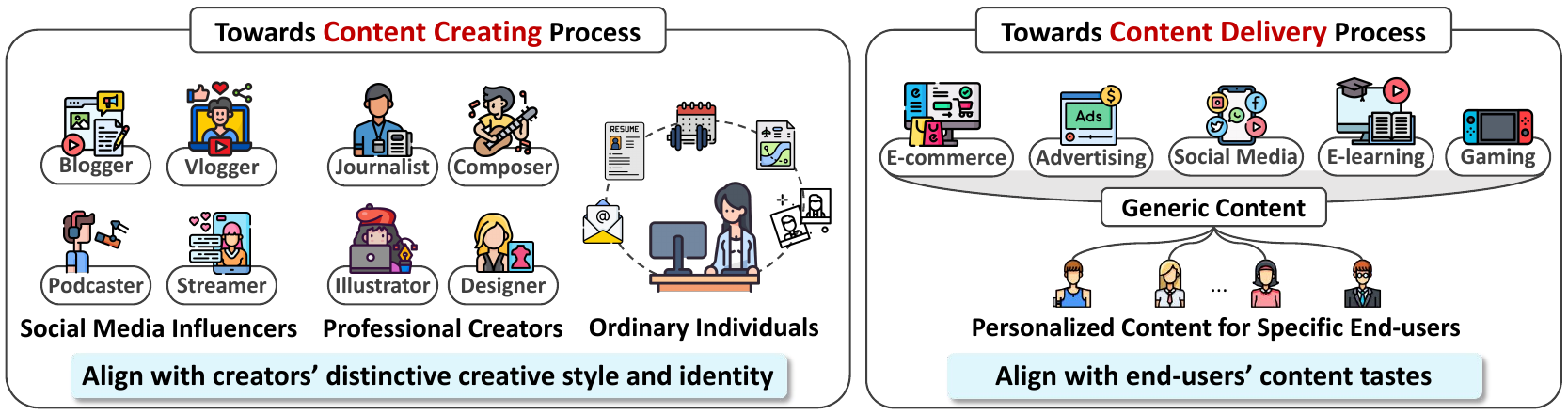}
\caption{Applications of personalized generation.} 
\label{fig:application}
\end{figure*}

\subsubsection{User behaviors}
Based on user interactions, generative models can infer user preferences to tailor robotic behaviors.

\paragraph{Robotics}
Several studies have investigated inferring user preferences from historical trajectories and associated human feedback, enabling personalized robotic decision-making. Specifically, \citet{poddar2024personalizing} proposed Variational Preference Learning, integrating variational inference into RLHF to model diverse user preferences and enable active learning. \citet{hwang2024promptable} introduced ``Promptable Behaviors'', using Multi-Objective Reinforcement Learning to dynamically customize policies via human demonstrations, trajectory feedback, and language instructions. \citet{cui2024board} presented a framework with a RAG memory module to match individual users’ preferences and driving styles.

\subsubsection{User Documents}
User-created documents, such as comments, reviews, and captions can be used to infer their writing style and preferences for personalization.

\paragraph{Caption/Comment Generation}
Several studies utilize user-created captions and comments to develop personalized captioning and comment systems \cite{shuster2019engaging, long2020cross, zhang2020learning, geng2022improving}. To effectively model user preferences, \citet{shin2018customized} engages users by soliciting answers to targeted questions during generation. \citet{lin2024personalized} and \citet{wu2024understanding} utilize users' historical comments for personalized video comment generation.
SimTube~\cite{hung2024simtube} incorporates user responses and defined personas to support user-steerable video comment generation.

In addition, PV-LLM~\cite{lin2024personalized} fine-tunes MLLMs using user-written comments, while PVCG~\cite{wu2024understanding} learns a unique identifier for each user to enhance personalization.

\subsubsection{Personalized subjects}
In some cases, users may provide specific subject images, such as photos of their friends, for personalized visual question answering (e.g., \emph{“Give me a birthday gift list for my friend Peter.”}).

\paragraph{Cross-modal Dialogue Systems}
Given user-specific subject images and queries, the systems are expected to identify these subjects and infer user intents for personalized responses. Existing studies in this domain mainly fall into two groups:
\begin{itemize}[leftmargin=*,itemsep=2pt,topsep=2pt,parsep=0pt]
    \item  Memory-based methods store user-specific subjects and activate or retrieve them as needed. For example, CSMN~\cite{chunseong2017attend, park2018towards} stores image memory, user context memory, and word output memory, and the model generates personalized captions based on memory features. \citet{hao2024remember} and \citet{seifi2025personalization} leverage RAG techniques to personalize pretrained MLLMs, allowing LLMs to update their supported concepts without requiring additional training. 
    PLVM~\cite{pham2024personalized} proposes a pre-trained Aligner to align referential concepts with the queried images. 
    In addition, R2P~\cite{das2025training} introduces a retrieval-reasoning paradigm to identify personal concepts in a training-free setting.
    
    \item Optimization-based methods inject personalized information into the generation process via specific modules. For example, PerVL~\cite{cohen2022my}, Yo'LLaVA~\cite{nguyen2024yollava}, and MC-LLaVA~\cite{an2024mc} learn to encode a personalized subject into a set of latent tokens based on several provided subject images. MyVLM~\cite{alaluf2025myvlm} introduces learnable heads, each dedicated to recognizing a single user-specific subject. 
    CaT~\cite{an2025concept} proposes a comprehensive data synthesis pipeline to enhance the personalization capabilities of MLLMs. PVChat~\cite{shi2025pvchat} empowers LLMs with personalization capabilities for subject-aware question answering with only one-shot learning.
    In addition, several studies have explored learning unique user embeddings \cite{long2020cross, zhang2020learning, xiong2020towards} or performing prefix tuning \cite{wang2023user} for personalization. 
\end{itemize}

\subsubsection{Evaluation Metrics}
To quantify the degree of personalization in text generation tasks such as personal assistant and comment generation, existing studies commonly employ  (1) \textit{term-matching metrics}, such as ROUGE, BLEU, Meteor, CIDEr~\cite{cohen2022my, alaluf2025myvlm, nguyen2024yollava, pi2024personalized, an2024mc}; (2) \textit{semantic matching metrics}, such as CLIPScore~\cite{cohen2022my}; (3)\textit{ recall, precision and F1-score} to validate whether the user-specific concept appears in the generated caption~\cite{cohen2022my, alaluf2025myvlm, nguyen2024yollava, pi2024personalized, an2024mc}; (4) \textit{human evaluation} to determine alignment with ground truths in terms of emotion, style, and relevance.

To measure whether agents align with diverse human preferences,  studies employ success rate, success weighted by path length (SPL), distance to goal, and episode length~\cite{poddar2024personalizing, hwang2024promptable}. Human evaluation is also an effective validation method for determining whether an agent has successfully completed a personalized task~\cite{hwang2024promptable, cui2024board}.

\section{Applications}
\label{sec:application}

The prior section has highlighted the success of PGen across various modalities, demonstrating its potential to enhance user engagement and enrich experiences across diverse domains. As illustrated in Figure~\ref{fig:application}, applications of PGen can be categorized based on the stages of content personalization:
1) towards content creation process, which provides personalized tools and services for content creators of all levels to maintain their unique creative style while streamlining the creative workflows; and 2) towards content delivery process, which delivers multimodal content in a personalized manner tailored to individual preferences of end users.

\subsection{Towards Content Creation Process}
Generative models have transformed the realm of content creation, pushing the boundaries of productivity and creativity. By incorporating personalization techniques, PGen can further empower content creators across various domains, allowing them to achieve higher efficiency while preserving their distinctive creative style and identity. 

For social media influencers, such as bloggers, vloggers, and podcasters, PGen can analyze their past content to offer tailored suggestions for compelling headlines~\cite{fang2024generate} and introductions, or even generate new content that aligns seamlessly with their established brand. This not only streamlines the creative process but also maintains their unique style, fostering deeper connections with their audiences.

For professional creators, such as journalists, designers, illustrators, and music composers, personal style serves as their creative fingerprint, crucial for building reputation and recognition within competitive creative industries. By analyzing creators' previous work, PGen can identify and adapt to their unique stylistic traits to provide tailored ideas, drafts, or modifications, striking a perfect balance between personal style and external demands.

Ordinary individuals can also benefit from PGen for routine tasks, such as personalized email drafting, resume creation, travel planning, workout scheduling, and portrait generation.

\subsection{Towards Content Delivery Process}
In the era of information overload, personalized content delivery is becoming increasingly essential for helping individuals navigate through vast amounts of multimodal content on the internet. By integrating PGen into the content delivery process, generic content can be adapted into diverse, personalized formats to engage different audiences, catering to their unique tastes and content needs. Below are representative application scenarios of PGen for personalized content delivery:

\textbf{Marketing and Advertising.} PGen can assist organizations and brands in creating targeted marketing strategies and dynamic advertisements that resonate deeply with specific audiences, ultimately driving higher click-through and conversion rates.

\textbf{Retail and E-commerce.} Through personalized product descriptions and images, customized manuals, and virtual try-ons, PGen empowers retailers to attract consumers and enhance engagement, delivering unique and tailored shopping experiences. 

\textbf{Entertainment and Media.} On digital content platforms such as Flipboard, Twitter, Netflix, and YouTube, personalized content plays a crucial role in attracting and retaining users. Examples include personalized news, posts, movie posters, video thumbnails, and other tailored media assets that can enhance user loyalty to the platform.

\textbf{Education and E-learning.} Generative models have shown significant promise in education, exemplified by platforms like Google Learn About~\footnote{https://learning.google.com/experiments/learn-about.}. PGen can further enhance personalized educational experiences by offering customized learning roadmaps and materials, dynamically adapting to individual learning styles, goals, and progress.

\textbf{Gaming.} Integrating PGen into the gaming industries enables the creation of dynamic storylines, customized tasks, scalable difficulty levels, and interactive characters that adapt to players' preferences and behaviors, fostering more immersive and engaging gaming experiences.

\textbf{Personalized AI Assistant.} PGen can be incorporated into AI assistants to provide specialized support, such as legal assistance, medical advice, and financial guidance, ensuring precision and user-specific customization.

\section{Open Problems}
\label{sec:open_problems}

Despite the significant progress made by PGen, several key challenges remain to be addressed. 
\subsection{Technical Challenges}
\begin{itemize}[leftmargin=*,itemsep=2pt,topsep=2pt,parsep=0pt]
\item \emph{\textbf{Scalability and Efficiency}}. 
PGen relies on large generative models for content personalization, which often require extensive resource costs, limiting their deployment to real-time, large-scale user scenarios. Developing scalable and efficient algorithms for PGen remains a critical direction for future research~\cite{yang2024efficient}. 

\item \emph{\textbf{Deliberative Reasoning for PGen}}. In certain personalized scenarios that prioritize content quality over temporal efficiency -- such as digital advertising, where advertisers typically serve only several ads per user each day -- inference scaling presents significant opportunities for enhancing user satisfaction. Prior work mainly focuses on multi-turn refinement to progressively enhance content relevance and personalization~\cite{nabati2024personalized}. Inspired by the great success of LLM reasoning~\cite{guan2024deliberative,guo2025deepseek,peng2025lmm}, 
deliberative reasoning may emphasize extensive logical and contextual reasoning beforehand, enabling a thorough analysis of user preferences to drive more effective content personalization~\cite{fang2025large}.

\item \emph{\textbf{Evolving User Preference}}. 
As explored in traditional RecSys~\cite{wang2023causal}, recognizing dynamic preferences from user behaviors is crucial to enhancing personalization. Adapting PGen to track and respond to user preference shifts remains a key research direction.

\item \emph{\textbf{Mitigating Filter Bubbles}}. PGen may inadvertently reinforce users’ existing preferences or beliefs and contribute to polarization~\cite{pmlr-v239-lazovich23a}, which has been widely studied in recommender systems. Strategies such as diversity enhancement~\cite{chen2018fast} and user-controllable inference~\cite{wang2022user} offer valuable insights for mitigating this in PGen. Moreover, multi-agent generation introduces diverse perspectives through distinct generative personas, providing a promising solution within generative settings~\cite{zhang2024see}.

\item \emph{\textbf{User Data Management}}. As the cornerstone of PGen, user data management is essential for collecting, storing, and structuring user-specific data, as well as enabling user-controllable personalization. Existing approaches primarily adopt either an external user memory module combined with RAG to manage user data~\cite{mysore2024pearlpersonalizinglargelanguage} or embed user-specific information directly into model parameters to achieve personalization~\cite{ruiz2023dreambooth}. However, developing a more effective and efficient strategy for lifelong user data management remains an open problem.

\item \emph{\textbf{Multi-modal Personalization}}.
Existing PGen research mainly focuses on single-modality generation, while multi-modal personalization remains underexplored, such as personalized social media posts that integrate both image and text. This challenge requires high-quality, instruction-aligned, and personalized output while ensuring consistency across multiple modalities.
Recently, the emergence of unified multimodal models, such as Show-o~\cite{showo}, Transfusion~\cite{zhou2024transfusion}, Janus-Pro~\cite{chen2025janus}, BAGEL~\cite{deng2025emerging}, and GPT-4o~\cite{openai2024gpt4o}, which provide shared architectures for personalized multimodal generation and open opportunities for transferring personalization techniques across modalities by leveraging insights from single-modality methods.

\item \emph{\textbf{Synergy Between Generation and Retrieval}}.
Traditional personalized content delivery systems primarily focus on retrieval-based methods like RecSys. However, existing content may not fully meet users' content needs. Integrating PGen with retrieval-based approaches holds great promise for building more powerful personalized content delivery systems~\cite{wang2023generative}.
\end{itemize}

\subsection{Benchmarks and Metrics}
A fundamental challenge in PGen is establishing robust metrics and benchmark datasets. Current evaluation methods primarily rely on traditional generation metrics (e.g., BLEU for text, CLIP-I for images), which do not fully capture how well the generated content aligns with user preferences. Future research should focus on developing more effective metrics for evaluating personalization.

\subsection{Trustworthiness}
Ensuring the trustworthiness of PGen is critical for fostering user confidence and promoting responsible deployment. Key considerations include:
\begin{itemize}[leftmargin=*,itemsep=2pt,topsep=2pt,parsep=0pt]
    \item \emph{\textbf{Privacy}}. PGen relies on user-specific data for content personalization, raising concerns about privacy issues. Striking a balance between effective personalization and robust privacy protection is crucial for advancing this field.
    Prior approaches, such as on-device personalization~\cite{wang2024mememo,cho2024hollowed,bang-etal-2024-crayon}, federated learning~\cite{jiang2024personalized,zhang2024fed}, differential privacy~\cite{flemings-etal-2024-differentially}, and adversarial perturbation~\cite{van2023anti}, offer promising strategies for achieving this balance, enabling personalization while safeguarding user privacy.

    \item \emph{\textbf{Fairness and Bias}}. PGen may unintentionally reinforce biases and stereotypes present in training data, leading to skewed or discriminatory outcomes~\cite{wan-etal-2023-personalized}. 
    Extensive studies have explored debiasing strategies for generative models, including context steering~\cite{pandey2024cos}, structured prompt~\cite{furniturewala-etal-2024-thinking,gallegos-etal-2025-self}, and causal-guided biased sample identification~\cite{sun-etal-2024-causal}, which offer valuable insights for mitigating potential biases in PGen.

    \item \emph{\textbf{Safety}}. Establishing transparent governance protocols, reliable moderation mechanisms, and explainable generation processes is crucial to maintaining user trust and upholding safety standards.
\end{itemize}

\section{Conclusion}
\label{sec:conclusion}

In this work, we present the first comprehensive survey on PGen across multiple modalities, offering an in-depth review of recent advancements and emerging trends in the field. To unify existing research, we introduce a holistic framework that formalizes diverse user-specific data, core objectives, and general workflows for PGen, providing a structured foundation for future developments. We then propose a multi-level taxonomy to categorize PGen methods based on modality, user inputs, and specific tasks. Additionally, we summarize the commonly used datasets and evaluation metrics for each modality. Beyond technical aspects, we explore PGen's potential applications in both content creation and delivery, underscoring its significant economic and research value. Finally, we identify key research challenges that remain to be addressed. As a rapidly evolving field, PGen holds great potential to revolutionize the online content ecosystem, enabling more tailored and engaging user experiences. By unifying PGen research across multiple modalities, this survey serves as a valuable resource for fostering cross-modal knowledge sharing and collaboration in this field, contributing to a more personalized digital landscape.

\section*{Acknowledgments}

This work is supported in part by the National Natural Science Foundation of China (62272437 and U24B20180), in part by Center for Intelligent Information Retrieval, in part by NSF grant number 2143434, and in part by an award from Google. Any opinions, findings, and conclusions or recommendations expressed in this material are those of the authors and do not necessarily reflect those of the sponsor.

\section*{Limitations}
In this paper, we provide a comprehensive survey of personalized generation. However, the rapid evolution of this field makes it challenging to encompass all research efforts, as new methods, datasets, and evaluation metrics continue to emerge, requiring continuous updates to our taxonomy. Furthermore, the development of more effective and universally accepted benchmarks within different modalities remains an ongoing challenge.

\bibliography{custom}

\end{document}